\documentclass{aa}
\usepackage{natbib}

\usepackage{twoopt}
\usepackage{graphicx}
\usepackage{amsmath}
\usepackage{txfonts}

\usepackage[para,online,flushleft]{threeparttable}

\usepackage[hidelinks]{hyperref}

\usepackage{url}
\bibpunct{(}{)}{;}{a}{}{,}

\newcommand{\teff}{\ensuremath{T_{\mathrm{eff}}}}
\newcommand{\logg}{\ensuremath{\log g}}
\newcommand{\vhelio}{\ensuremath{V_{\mathrm{HELIO}}}}

\newcommand{\kms}{\ensuremath{\mathrm{km\,s}^{-1}}}
\newcommand{\masyr}{\ensuremath{\mathrm{mas\,yr}^{-1}}}
\newcommand{\Msun}{\ensuremath{\mathrm{M}_\odot}}
\newcommand{\Msunyr}{\ensuremath{\mathrm{M}_\odot\,\mathrm{yr}^{-1}}}

\newcommand{\feh}{\ensuremath{[\mathrm{Fe}/\mathrm{H}]}}

\newcommand{\mgfe}{\ensuremath{[\mathrm{Mg}/\mathrm{Fe}]}}

\begin{document}

\title{The chemical evolution of the LMC from APOGEE abundances: a hybrid two-infall OMEGA+ analysis}

\author{
Erik Gasteiger\inst{1,2},
Viola Heged\H{u}s\inst{1,2,3},
Szabolcs M{\'e}sz{\'a}ros\inst{2,3,4}
}

\institute{
ELTE E\"otv\"os Lor\'and University, Institute of Physics and Astronomy, P\'azm\'any P\'eter s\'et\'any 1/A, 1117 Budapest, Hungary
\and
ELTE E\"otv\"os Lor\'and University, Gothard Astrophysical Observatory, 9700 Szombathely, Szent Imre H. st. 112, Hungary
\and
MTA-ELTE Lend{\"u}let "Momentum" Milky Way Research Group, Szent Imre h. u. 112., Szombathely H-9700, Hungary
\and
HUN-REN CSFK, Konkoly Observatory, Konkoly Thege Mikl\'os \'ut 15-17, Budapest, 1121, Hungary
}

\abstract
{The Large Magellanic Cloud (LMC) provides resolved stellar abundances and star-formation histories that can be used together to test models of dwarf-galaxy chemical evolution.}
{We investigated if the non-monotonic APOGEE $[\mathrm{Mg}/\mathrm{Fe}]$--$[\mathrm{Fe}/\mathrm{H}]$ sequence can be explained by a hybrid history that combines a second gas-infall episode with a temporary increase in star-formation efficiency (SFE).}
{We selected 4933 LMC red-giant-branch stars from APOGEE-2 DR17. We fitted an adaptively binned median Mg sequence with the two-zone open-box code \textsc{OMEGA+}. The formal fitting objective contained only the Mg sequence. The metallicity distribution function (MDF), one-dimensional abundance distributions, star-formation history (SFH), present-day mass partition, SFR, and supernova (SN) rate were retained as posterior diagnostics.}
{The best-fitting model reproduces the Mg sequence with an RMS residual of 0.01~dex and also gives a close posterior match to O. It contains two delayed-tau infall episodes, with the second beginning at $t_2=10.39$~Gyr and with integrated mass ratio $M_2/M_1=0.159$, followed by a temporary Gaussian SFE enhancement centred at $t=11.91$~Gyr. The second infall is represented by a mixture of enriched and primordial gas. The model reproduces the timing of the late SFH enhancement and yields a present SFR within the adopted comparison range. }
{A hybrid second-infall and SFE-enhancement history is a plausible scenario for reproducing the observed LMC Mg median sequence. Our fit explains the LMC evolution fairly well, but the remaining small MDF, mass, and SN-rate discrepancies suggest that a selection-function-aware treatment of the stellar distributions is required for a stricter comparison.}


\titlerunning{Chemical evolution of the LMC}
\authorrunning{Gasteiger et al.}
\maketitle

\section{Introduction}
\label{sec:intro}

Nearby dwarf galaxies offer an excellent opportunity for testing how gas accretion, star formation, material outflows, and supernova enrichment shape chemical evolution. The Large Magellanic Cloud (LMC) is a key benchmark because its proximity enables resolved constraints on both its star-formation history and detailed stellar abundances, allowing direct confrontation with global chemical evolution models \citep{harris2009}.

Large, homogeneous spectroscopic surveys now provide the multi-element chemistry required for such tests. APOGEE-2, within SDSS-IV \citep{esienstein2011,blanton2017}, delivered high-resolution near-infrared spectra and calibrated abundances for LMC red giants \citep{majewski2017,abdurrouf2022}. Combined with Gaia astrometry \citep{gaia_dr3}, these data enable clean membership selection and precise constraints in abundance planes such as \mgfe--\feh, which track the relative timing of core-collapse and Type~Ia enrichment. Interpreting these patterns commonly relies on galactic chemical evolution (GCE) models that track gas mass, star formation, and abundances under parameterised inflow, outflow, and star-formation efficiency \citep{pagel1998}. Modern implementations extend this framework with flexible gas cycling and delayed recycling, enabling direct comparison of enrichment tracks with time-dependent mass flows \citep{cote2017,cote2018}.

A central difficulty is non-uniqueness. Different combinations of inflow history, ${\rm SFE}(t)$, and outflow efficiency can reproduce similar \mgfe--\feh\ trends while predicting different present-day gas and stellar masses and supernova rates. Recent one-zone interpretations of the APOGEE LMC abundances reproduce the late-time $\alpha$-element "bump" by invoking a late enhancement of SFE, without requiring a distinct additional gas-infall episode \citep{nidever2020,hasselquist2021}. In contrast, external accretion has been motivated by the age--metallicity relation, where an apparent metallicity dip could be explained by the addition of $\sim10^{9}\,M_{\odot}$ of metal-poor gas \citep{bekki2012}. For this reason, the median abundance track alone is not a complete validation of a chemical-evolution scenario. Independent diagnostics such as the SFH, cumulative SFH, metallicity distribution function (MDF), abundance-ratio distributions, and present-day mass components are needed to test whether the same model also predicts a plausible stellar population and time-domain evolution.

Resolved SFH reconstructions support a late increase in star formation in the last few Gyr \citep{harris2009,weisz2013,meschin2014}, while interaction models predict that metal-poor gas associated with the Magellanic system can be transferred to the LMC at late times, providing a plausible channel for discrete infall episodes \citep{bekki2007}. New APOGEE-based observational analyses further show that many radial abundance gradients and age--abundance trends change their behaviour around $\sim2$~Gyr ago, consistent with a major event affecting LMC chemical evolution on that timescale \citep{povick2025}. Finally, recent one-zone LMC chemical evolution models developed to predict short-lived radionuclides test multiple SFH prescriptions and select a scenario with a mild early star-formation rate and a recent burst that simultaneously matches present-day constraints and $\alpha$-element abundance patterns \citep{vasini2023}. These studies show that similar late-time chemical signatures can be reproduced by different physical drivers, and that present-day mass estimates, SFR and SN rates help discriminate between them.

In this paper, we used APOGEE-2 (SDSS-IV DR17) abundances of LMC RGB stars and modelled the evolution with the two-zone open-box code \textsc{OMEGA+} \citep{cote2018}. We fitted the APOGEE \mgfe--\feh\ median sequence with a hybrid model in which a late mixed-composition infall episode and a time-dependent SFE act together. We then evaluated the fitted history against the APOGEE MDF and abundance-ratio distributions, the SFHs of \citet{harris2009} and \citet{weisz2013}, the present-day mass components, SFR, and SN rate.
This paper is structured as follows: Sect.~\ref{sec:data_selection} describes the sample and membership selection, Sect.~\ref{sec:gce_model} outlines the GCE framework and model setup, Sect.~\ref{sec:results} presents the results and their implications, and Sect.~\ref{sec:conclusions} summarises the conclusions.

\section{Target and data selection}
\label{sec:data_selection}

\begin{figure*}
\centering
\includegraphics[width=7.1in,angle=0]{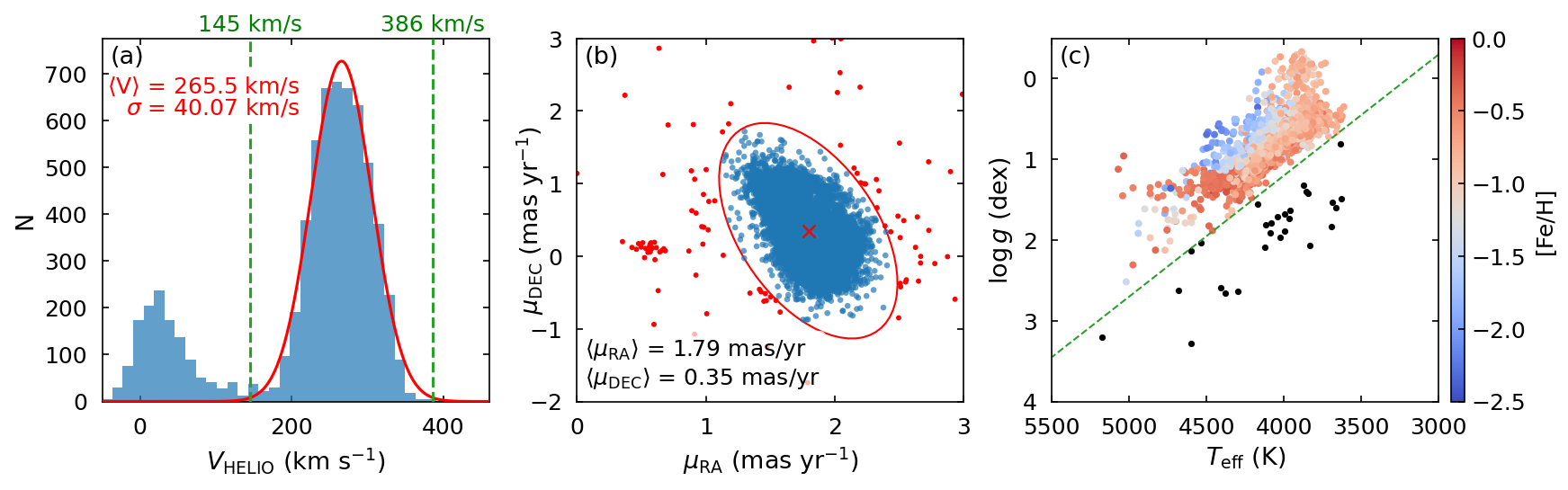}
\caption{Sequential LMC membership selection for the APOGEE--Gaia sample.
(a) Distribution of heliocentric radial velocities $\vhelio$, showing a bimodal behaviour associated with the Milky Way and LMC components. The red curve is a Gaussian fit to the LMC component with $\langle V\rangle=265.5~\kms$ and $\sigma=40.07~\kms$. The vertical green dashed lines mark the adopted kinematic membership window $\langle V\rangle\pm3\sigma$ (145--386~\kms).
(b) Gaia DR3 proper-motion plane $(\mu_{\rm RA},\mu_{\rm DEC})$: blue points are stars retained by the proper-motion selection, red points are rejected outliers, the red ellipse indicates the adopted selection region, and the red cross marks the mean proper motion of the LMC.
(c) Kiel diagram ($\logg$--$\teff$) for the selected LMC RGB stars. The green dashed line shows the linear cut used to isolate the red giant branch. Points are colour-coded by metallicity $\feh$, and black points indicate stars excluded by this criterion.}
\label{fig:fig1}
\end{figure*}

We constructed a homogeneous LMC sample from the APOGEE-2 near-infrared ($1.51$--$1.70~\mu$m), high-resolution ($R=22\,500$) spectroscopic survey \citep{majewski2017}. The adopted calibrated stellar parameters (\teff, \logg), heliocentric radial velocities (RV; \vhelio), and elemental abundances were taken from the SDSS-IV DR17 release \citep{abdurrouf2022}, as derived by the APOGEE Stellar Parameters and Chemical Abundances Pipeline (ASPCAP; \citealt{garciaperez2016}). We complemented these data with Gaia Data Release 3 (DR3) astrometry, using proper motions (PM; $\mu_{\rm RA},\mu_{\rm DEC}$) for membership selection \citep{gaia_dr3}. Following standard extragalactic target selection practice, our strategy closely follows the approach used for APOGEE Magellanic targets by \citet{nidever2020,hasselquist2021}. We aimed for a kinematically clean RGB sample with high abundance precision and minimal foreground contamination, suitable for chemical evolution modelling.

We excluded stars marked with the \texttt{STAR\_BAD} flag because their measurements are affected by known data-quality problems (e.g., problematic pixels, spectroscopic binaries, or unreliable atmospheric parameters). We required $\mathrm{SNR}>70$ to ensure robust abundance measurements and restricted to stars fainter than $H=12.3$ to retain Magellanic targets. The LMC spatial candidate sample was set by selecting stars within $30^\circ$ of the LMC centre at $\alpha_{\mathrm{LMC}} = 80.893860^\circ$ and $\delta_{\mathrm{LMC}}=-69.756126^\circ$ \citep{nidever2020}.

The RV distribution shows two components in Fig.~\ref{fig:fig1}: a low-velocity Milky Way foreground and a high-velocity LMC peak. We isolated the LMC component by fitting a Gaussian to the high-velocity peak after identifying the local minimum separating it from the foreground. The fit yields $\langle V\rangle=265.5~\kms$ and $\sigma=40.07~\kms$, and stars were retained within $\langle V\rangle\pm3\sigma$.

We further refined membership in the Gaia DR3 PM plane (Fig.~\ref{fig:fig1}). To define an objective and reproducible selection region, an ellipse was fitted to the main clump (implemented with \texttt{EllipticEnvelope} in \texttt{scikit-learn}). The obtained centroid is $\langle\mu_{\mathrm{RA}}\rangle=1.79~\masyr$ and $\langle\mu_{\mathrm{DEC}}\rangle=0.35~\masyr$, with standard deviations along the principal axes of $\sigma_a=0.51~\masyr$ and $\sigma_b=0.20~\masyr$ and position angle $\Theta=105^\circ$. Stars were selected within the corresponding $3\sigma$ ellipse.

We restricted the sample to RGB stars using ASPCAP parameters, adopting $3600~\mathrm{K}<\teff<5200~\mathrm{K}$ and $\logg<3.4~\mathrm{dex}$ \citep{nidever2020,majewski2017}. A linear cut was additionally applied in the Kiel diagram (Fig.~\ref{fig:fig1}) to remove residual young and main-sequence contaminants. The final LMC sample contains $4933$ RGB stars.

In the following sections, we use the abundance trends of this quality-controlled APOGEE--Gaia RGB sample to constrain our chemical evolution model of the LMC.

\section{Chemical evolution model}
\label{sec:gce_model}

\subsection{Model overview}

We modelled the chemical evolution of the LMC with \textsc{OMEGA+}, an open-source \texttt{Python} code within the JINA--NuGrid \textsc{JINAPyCEE} ecosystem \citep{cote2018}. \textsc{OMEGA+} is a two-zone, open-box model in which the star-forming cold gas is embedded in a hot circumgalactic medium (CGM) reservoir, enabling two-step gas exchange and delayed recycling \citep{cote2018}. The model returns time-dependent mass components, star-formation rate, inflow/outflow rates, SN rates, and elemental abundances. The primary observational constraint was the \mgfe--\feh\ track. The standard governing equations are given by \citet{cote2017,cote2018}.

\subsection{Two-infall accretion history}

The hybrid setup is motivated by models of the recent LMC--SMC interaction, which can transfer gas and enhance star formation \citep{bekki2007,besla2012,bekki2012}, together with observational evidence for late star formation and abundance changes in the LMC \citep{harris2009,nidever2020,hasselquist2021}. In our model, the mixed second infall represents the late gas supply, while the Gaussian SFE enhancement represents the associated increase in star-formation activity.

We adopted two independently normalised delayed-tau infall episodes. This form provides a gradual early gas supply and a distinct late event without fixing the ratio of their instantaneous amplitudes. For episode $j$,

\begin{equation}
\dot{M}_{\mathrm{in},j}(t)=A_j x_j\exp(-x_j)\,\Theta(t-t_j),
\qquad x_j=\frac{t-t_j}{\tau_j},
\label{eq:two_infall}
\end{equation}

where $t_1=0$, $t_2$ is the onset of the second episode, and $\tau_1$ and $\tau_2$ are the characteristic widths. The coefficient $A_j$ sets the rate normalisation and therefore has units of mass per unit time. For a calculation ending at $t_{\rm end}$, the integrated mass of episode $j$ is

\begin{equation}
\begin{split}
M_j &=\int_{t_j}^{t_{\rm end}}\dot{M}_{\mathrm{in},j}(t)\,\mathrm{d}t \\
M_j &=A_j\tau_j\left[1-(1+X_j)\exp(-X_j)\right],
\qquad X_j=\frac{t_{\rm end}-t_j}{\tau_j}.
\end{split}
\label{eq:infall_mass}
\end{equation}

We fitted the finite-horizon mass ratio $M_2/M_1$ directly and derived $A_1$ and $A_2$ from the corresponding integrated masses. This ratio determines the relative total mass supplied by the two infall episodes over the model time. It is more informative than a ratio of peak rates when the episodes have different widths. The absolute scale was applied after the chemical calculation so that the final baryonic mass was $3.5\times10^9\,\Msun$. We adopted this value from the upper limit to the total mass within a radius of 4~kpc reported by \citet{kim1998}. This value serves as the model normalisation. It is not a direct measurement of the LMC baryonic mass.

The first infall was assumed primordial ($Z=0$). For the second infall, we used a mixture of pre-enriched and primordial gas. Here, pre-enriched means that the gas has a non-zero metallicity. In our model, 85\% of the infalling mass has $\feh=-0.65$ and $\mgfe=0.30$, while the remaining 15\% is primordial. The abundances of Mg, and Fe were from the solar abundance table of \citet{grevesse2007}. For all other metals, we adopted the relative isotopic proportions of the GN93 solar mixture supplied with \textsc{JINAPyCEE} \citep{grevesse1993}. This provides the complete elemental composition required by the model. After dilution by the primordial component, the accreted gas has $\feh=-0.72$, while $\mgfe$ remains 0.30. The composition and mixing fraction are adopted model parameters and are not direct measurements of the transferred gas.

\subsection{Fixed and free parameters}

The SFR was computed as $\dot{M}_\star=[\epsilon_\star(t)/\tau_\star]M_{\rm g}$, with $\tau_\star=0.1t_{\rm dyn}$ as implemented in \textsc{OMEGA+} \citep{cote2018}. Here $t_{\rm dyn}$ is the dynamical time of the virialised system. The time-dependent efficiency was written as $\epsilon_\star(t)=\epsilon_{\star,0}S(t)$, where

\begin{equation}
S(t)=1+(A_{\rm peak}-1)\exp\!\left[-\frac{(t-t_{\rm SFE})^2}{2\sigma_{\rm SFE}^2}\right].
\label{eq:sfe_history}
\end{equation}

Here $A_{\rm peak}$ is the maximum SFE multiplier, so $S(t_{\rm SFE})=A_{\rm peak}$, and $\sigma_{\rm SFE}$ is the Gaussian width. We fitted the offset $t_{\rm SFE}-t_2$. Outflows followed the standard \textsc{OMEGA+} mass-loading equation $\dot{M}_{\rm g,out}=\eta\dot{M}_\star$. This links the wind rate to recent star formation because the stellar feedback that drives the wind is produced by newly formed massive stars.

SN~Ia enrichment used a power-law delay-time distribution (DTD), $\mathrm{DTD}(t)\propto t^{\beta_{\rm Ia}}$, with the slope and the time-integrated SN~Ia number per formed stellar mass included in the fit \citep{maoz2012}. We also fitted the IMF exponent in $\mathrm{d}N/\mathrm{d}M\propto M^{-\alpha_{\rm IMF}}$. CCSNe were produced by stars with $8\leq M/\Msun\leq30$, while direct collapse was assumed above $30\,\Msun$ \citep{matteucci2021}. Element production and mass return were calculated with the \textsc{SYGMA} yield tables provided within \textsc{NuPyCEE} \citep{cote2017,cote2018}. The Mg yields were multiplied by a fitted factor $f_{\rm Mg}$ to account for yield-set systematics \citep{francois2004,spitoni2021}.

The bounds in Table~\ref{tab:gce_params} were fixed before the formal fit. The limits on the late-event timing encompass the recent LMC star-formation enhancement and abundance changes found in observational and chemical-evolution studies \citep{harris2009,nidever2020,hasselquist2021,povick2025}. The IMF interval covers the Salpeter exponent and the steeper IMFs previously considered for the LMC \citep{salpeter1955b,bekki2012,vasini2023}, while the SN~Ia DTD limits encompass empirical slopes near $t^{-1}$ and the observed time-integrated SN~Ia production per unit formed stellar mass \citep{maoz2012}. The $f_{\rm Mg}$ interval includes the factor-of-seven empirical correction applied by \citet{francois2004} to Mg yields from stars with masses of $11$--$20\,\Msun$, although our multiplier is applied uniformly to the adopted Mg yield set. The remaining bounds were chosen from preliminary calculations to encompass physically valid parameter combinations that produced plausible Mg tracks. 

The integration used an adaptive grid beginning at 0.1~Myr. The earlier mechanism searches used a maximum step of 32.5~Myr, while the formal fit and the adopted model used a maximum step of 16.25~Myr. Table~\ref{tab:gce_params} lists the 13 fitted parameters of the adopted history.

\subsection{Fitting method}

The LMC RGB dataset contains 4933 stars. Of these, 4864 have finite Fe and Mg abundances inside the formal interval $-1.8\leq\feh<-0.3$. After sorting them by \feh, we defined a histogram containing 32 consecutive groups of 150 stars, leaving 64 stars at the metal-rich end. Because this remainder contained fewer than 75 stars, we added it to the preceding group, which then contained 214 stars. We subsequently merged neighbouring groups wherever their \feh\ width was smaller than 0.03~dex. The final target therefore contains 19 non-overlapping bins with 150--450 stars per bin. The observed value in each bin is the median \mgfe. For a bin containing $N$ stars, we generated 1000 bootstrap samples. Each sample was constructed by making $N$ independent random draws from the original $N$ stars and returning the selected star after every draw. Each bootstrap sample therefore contained $N$ entries, but individual stars could appear more than once while others could be absent. We calculated the median for every bootstrap sample and adopted the standard deviation of the resulting 1000 medians as the sampling uncertainty ($\sigma_{i,\mathrm{bootstrap}}$). In highly populated bins, this uncertainty can become very small, which would give those bins excessive weight in the fit. We therefore imposed a minimum uncertainty of 0.02~dex, comparable to the $0.021$--$0.033$~dex median abundance precision reported for high-SNR APOGEE LMC stars by \citet{nidever2020}. The 45 stars at $-2.3\leq\feh<-1.8$ were retained as a metal-poor diagnostic and did not enter the fit. For the model comparison, we applied the same \feh\ bin boundaries to the model simple stellar populations (SSPs) and calculated the median \mgfe\ in each bin.

\begin{equation}
	r_i=\frac{[\mathrm{Mg}/\mathrm{Fe}]_{i,\rm model}-[\mathrm{Mg}/\mathrm{Fe}]_{i,\rm obs}}
	{\max\!\left(\sigma_{i,\mathrm{bootstrap}},0.02~\mathrm{dex}\right)},
	\qquad \chi^2_{\rm Mg}=\sum_i r_i^2.
	\label{eq:mg_objective}
\end{equation}

O, Si, Ca, the MDF, the one-dimensional abundance distributions, the SFH, the mass components, the present SFR, and the SN rate did not affect $\chi^2_{\rm Mg}$. They were used as posterior diagnostics.

The starting parameter vector for the local fit was selected from a broad preliminary scan of the parameter ranges in Table~\ref{tab:gce_params}, performed for both the enriched and mixed late-inflow prescriptions. These searches identified models combining a mixed-composition second infall with a temporary SFE enhancement as the most promising prescription and informed the parameter bounds used in the formal fit. We then used \texttt{lmfit} \citep{lmfit} for a bounded least-squares fit of the final hybrid two-infall model, with all 13 parameters in Table~\ref{tab:gce_params} varied simultaneously. The late-inflow composition and infall-profile family were fixed during this fit. Only $f_{\rm Mg}$ was included among the fitted element-yield multipliers. O, Si, and Ca multipliers were calibrated separately and held fixed. 

After fixing the Mg-derived history, we calibrated O, Si, and Ca independently. We used the same \feh\ bin boundaries and bootstrap uncertainty procedure as for Mg. For each element, only its yield multiplier was varied, and a deterministic bounded scalar search minimised the median-track $\chi^2$. The tested intervals were $0.5\leq f_{\rm O}\leq8.0$, $1.0\leq f_{\rm Si}\leq4.5$, and $1.5\leq f_{\rm Ca}\leq6.0$. These posterior calibrations did not alter the history parameters or the Mg objective.

\begin{scriptsize}
\begin{table*}[!t]
\caption{Parameters of the hybrid two-infall model.}
\label{tab:gce_params}
\centering
\setlength{\tabcolsep}{5pt}
\renewcommand{\arraystretch}{1.05}
{
\begin{tabular}{llll}
\hline
Parameter & Meaning & Bounds & Best fit \\
\hline
$\tau_1$ [Gyr] & first delayed-tau width & 0.7--2.5 & 1.718 \\
$t_2$ [Gyr] & second-infall onset & 9.0--11.4 & 10.391 \\
$\tau_2$ [Gyr] & second delayed-tau width & 0.3--1.5 & 0.481 \\
$M_2/M_1$ & integrated infall-mass ratio & 0.06--0.20 & 0.159 \\
$\epsilon_{\star,0}$ & baseline SFE parameter & $8\times10^{-4}$--$3\times10^{-3}$ & $9.79\times10^{-4}$ \\
$t_{\rm SFE}-t_2$ [Gyr] & SFE-event delay & 0.3--1.8 & 1.515 \\
$A_{\rm peak}$ & Gaussian SFE factor & 10--32 & 17.511 \\
$\sigma_{\rm SFE}$ [Gyr] & Gaussian SFE width & 0.8--3.0 & 1.116 \\
$\eta$ & outflow mass loading & 1.5--6.5 & 5.184 \\
$\alpha_{\rm IMF}$ & IMF exponent & 2.25--2.78 & 2.684 \\
$\beta_{\rm Ia}$ & SN~Ia DTD exponent & $-1.3$--$-0.8$ & $-1.293$ \\
$N_{\rm Ia}/M_\star$ [$\Msun^{-1}$] & integrated SN~Ia number & $5\times10^{-4}$--$1.5\times10^{-3}$ & $7.53\times10^{-4}$ \\
$f_{\rm Mg}$ & Mg yield multiplier & 4.5--9.5 & 7.475 \\
\hline
\end{tabular}
}
\tablefoot{All 13 parameters were varied together in the formal fit. The bounds define the adopted physical search domain and are not covariance-matrix uncertainties.}
\end{table*}
\end{scriptsize}

The fitted SN~Ia DTD exponent, $\beta_{\rm Ia}=-1.293$, lies close to the lower bound of $-1.3$. This indicates that the Mg-only fit favours a steep DTD within the tested domain. Independent SN constraints would be needed to determine this parameter more robustly.

\section{Results and discussion}
\label{sec:results}

\subsection{Chemical abundance trends and model fit}
\label{subsec:chem}

\begin{figure*}
\centering
\includegraphics[width=\textwidth]{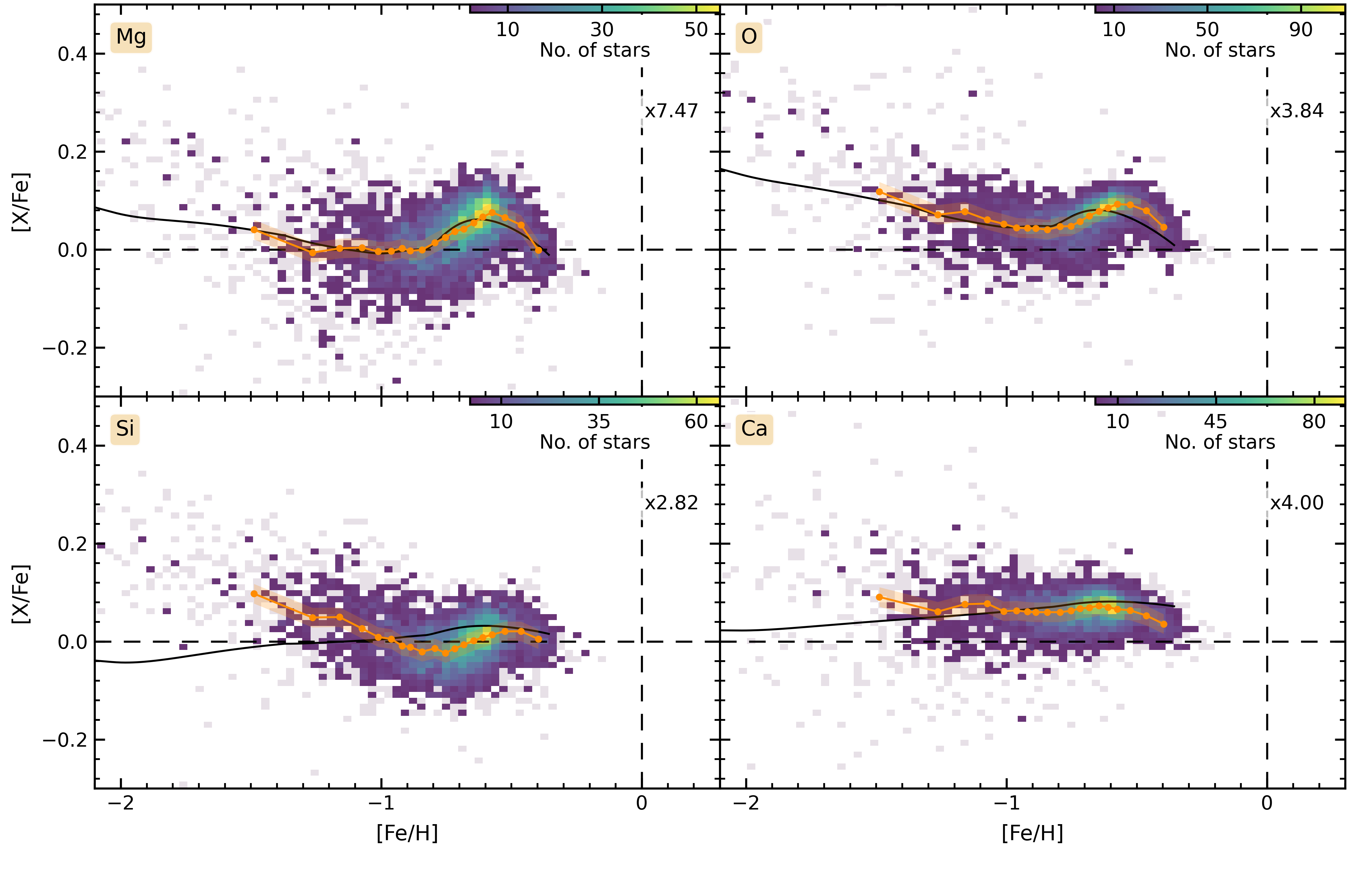}
\caption{Observed APOGEE LMC abundance distributions compared with the selected hybrid two-infall \textsc{OMEGA+} model. The density maps show the APOGEE RGB sample. Orange points show the adaptive-bin medians, and the orange bands show the uncertainties used in the fits, $\max(\sigma_{i,\mathrm{bootstrap}},0.02~\mathrm{dex})$. The black curve is the model track. The history was fitted to Mg only. O, Si, and Ca show posterior element-wise yield calibrations.}
\label{fig:fig2}
\end{figure*}

Figure~\ref{fig:fig2} compares the APOGEE LMC abundance distributions with the selected hybrid two-infall model. The Mg sequence is the formal fit. The quoted RMS therefore refers to these binned model medians. The black curves in Fig.~\ref{fig:fig2} show the fitted evolutionary tracks, which closely follow the corresponding binned model sequences over the fitted interval. O, Si, and Ca are posterior comparisons whose element-wise yield multipliers were not included in the formal history fit. The Mg sequence is reproduced over the full fitted interval with an RMS residual of 0.010~dex, which indicates excellent agreement. The adopted solution has $\chi^2_{\rm Mg}=4.99$. With 19 fitted bins and 13 fitted parameters, this corresponds to six nominal degrees of freedom and $\chi^2_\nu=0.83$. The observed bump reaches its maximum at $(\feh,\mgfe)=(-0.584,0.073)$, while the model bump reaches its maximum at $(-0.637,0.061)$, providing an agreement that is well within observational errors of the APOGEE measurements.

The posterior O sequence is also reproduced closely, with an RMS residual of 0.015~dex. Ca gives an acceptable match with an RMS of 0.019~dex. Si remains less satisfactory, with an RMS of 0.038~dex and a different track shape. The adopted yield multipliers are $f_{\rm Mg}=7.475$, $f_{\rm O}=3.836$, $f_{\rm Si}=2.823$, and $f_{\rm Ca}=4.001$. These factors correct the adopted yield set and should not be interpreted as direct measurements of stellar yields.

The fitted history requires two linked late-time changes. The second infall begins at $t_2=10.391$~Gyr and peaks at 10.872~Gyr. The fitted ratio is $M_2/M_1=0.159$, so the second episode supplies 13.7\% of the total accreted mass, or approximately $2.0\times10^9\,\Msun$ after the final mass normalisation. At the 13~Gyr endpoint the declining Gaussian tail gives $S(t)=11.2$. The Gaussian SFE enhancement peaks later, at $t_{\rm SFE}=11.906$~Gyr, where the SFE reaches $A_{\rm peak}=17.5$ times its baseline value ($\epsilon_{\star,0}=9.79\times10^{-4}$). The second infall peaks approximately 1~Gyr before the SFE enhancement. The gas reservoir is therefore replenished before star formation becomes most efficient, and the resulting increase in CCSN enrichment produces the Mg upturn. 

\begin{figure*}
\centering
\includegraphics[width=\textwidth]{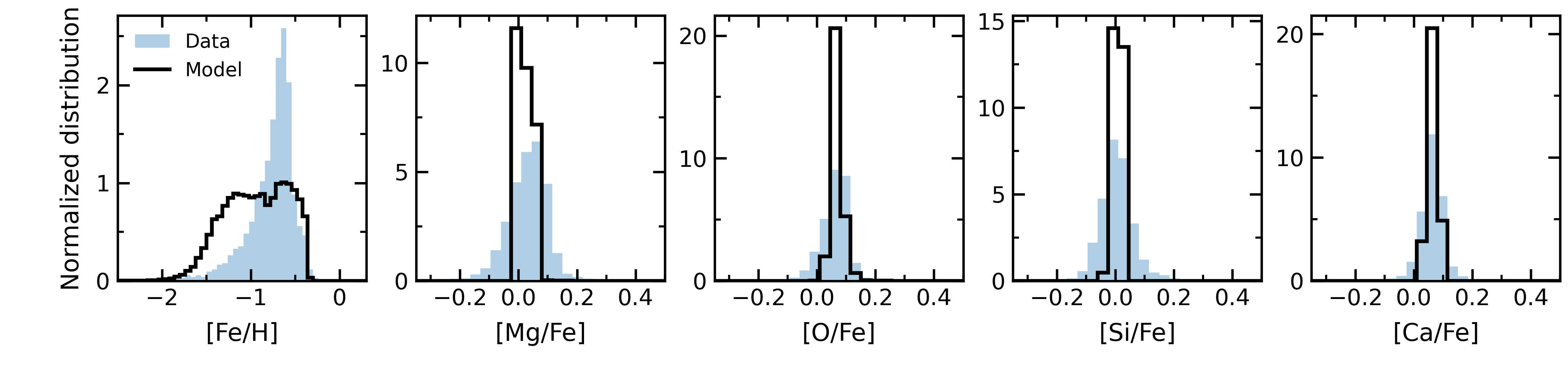}
\caption{Comparison of the APOGEE LMC RGB sample with the selected model in one-dimensional chemical distributions. Blue histograms show the observed sample and black step curves show the model distributions weighted by formed stellar mass. No RGB selection function was applied to the model. These distributions were not included in the fitting objective.}
\label{fig:fig3}
\end{figure*}

Figure~\ref{fig:fig3} compares the one-dimensional chemical distributions. The model and APOGEE MDFs reach their maximum in the same bin, centred at $\feh=-0.63$. The model reproduces the position of the principal MDF peak, although its peak is lower and its distribution is broader. Thus, the model predicts more metal-poor stars than what is observed in the LMC, while not producing enough metal-rich stars. The model Mg and O distributions show a much better agreement with the observations, as they peak at approximately the same abundance ratios as the APOGEE distributions (well within the observational errors of APOGEE \citep{abdurrouf2022}), although the model profiles are narrower and more strongly concentrated around their peaks.

The observed histogram contains APOGEE-selected RGB stars, whereas the model histogram is weighted by the formed stellar mass over all timesteps. The physical origin of the model's broad metal-poor tail is that it forms too much stellar mass while the gas remains at low metallicity. Its narrow Mg and O histograms arise because much of the modelled stellar mass forms during the late episode over a relatively small abundance range. The selection mismatch affects the exact size of these differences. The MDF and abundance histograms therefore provide useful qualitative checks. A quantitative comparison would require either a selection-corrected observed distribution or a forward selection model. Reducing the physical metal-poor excess would require less star formation before enrichment or faster early enrichment, which could be tested through a shorter first infall, a different early SFE, or time-dependent outflow in the future.

\subsection{Time evolution and present-day constraints}

\begin{figure}
\centering
\includegraphics[width=3.54in]{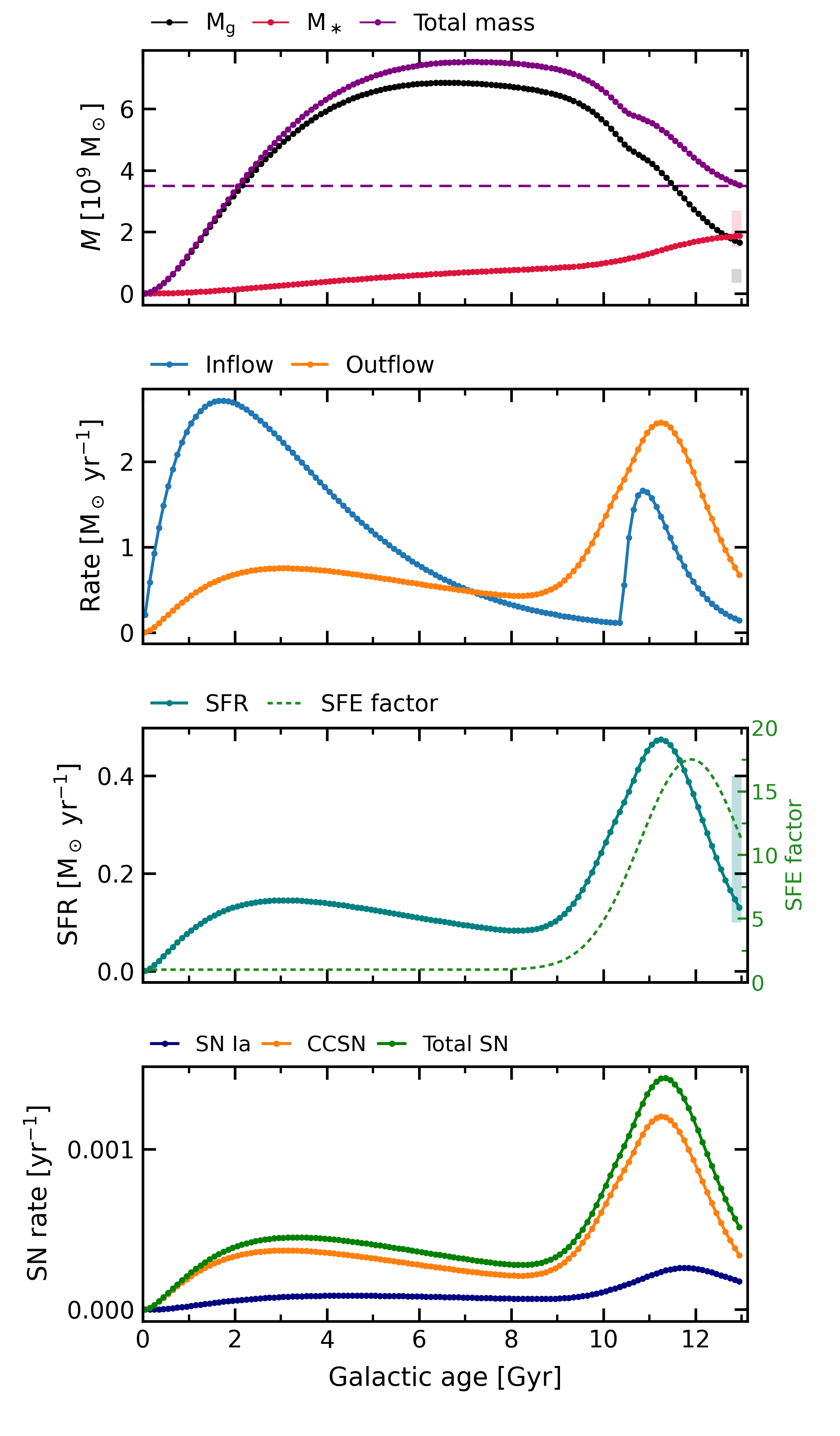}
\caption{Time evolution of the selected hybrid model. The panels show gas, surviving stellar, and total baryonic mass, inflow and outflow rates, SFR and the SFE factor, and SN~Ia, CCSN, and total SN rates. The dashed purple line marks the adopted final baryonic mass of $3.5\times10^9\,\Msun$. The green dashed curve in the SFR panel shows the SFE factor on the right axis. Markers at the present time indicate the neutral-gas estimate, the span of published stellar-mass estimates, and the adopted present-SFR comparison range \citep{kim1998,mcconnachie2012,vandermarel2002,eskew2012,skibba2012,harris2009}.}
\label{fig:fig4}
\end{figure}

Figure~\ref{fig:fig4} shows the mass reservoirs, gas flows, SFR, SFE factor, and SN rates. The final model scale was normalised to $M_{\rm g}+M_\star=3.5\times10^9\,\Msun$. The resulting present-day masses are $M_{\rm g}=1.62\times10^9\,\Msun$ and $M_\star=1.88\times10^9\,\Msun$. Published observation-based estimates of the LMC stellar mass span approximately $(1.5$--$2.7)\times10^9\,\Msun$ \citep{mcconnachie2012,vandermarel2002}. Spatially resolved SFH and infrared analyses give values of approximately $(1.9$--$2.0)\times10^9\,\Msun$ \citep{eskew2012,skibba2012}. The model stellar mass is therefore in excellent agreement with these estimates. The model gas reservoir is about three times higher than the neutral-gas estimate of $5.2\times10^8\,\Msun$ \citep{kim1998}. While the model gives a broadly plausible baryonic partition, it retains more gas than indicated by the adopted neutral-gas estimate.

The late infall peaks at $t=10.87$~Gyr and the SFR reaches $0.474\,\Msunyr$ at $t=11.25$~Gyr. \citet{harris2009} reported an average recent SFR of approximately $0.2\,\Msunyr$, with temporal variations at the factor-of-two level. We therefore adopted $0.1$--$0.4\,\Msunyr$ as a broad comparison range, rather than as a formal confidence interval. The model present SFR of $0.124\,\Msunyr$ lies within this range. Thus, the model produces both a prominent second infall and a late star-formation enhancement of approximately the required timing and strength.

The total model SN rate peaks at $1.44\times10^{-3}\,\mathrm{yr}^{-1}$ and reaches $5.15\times10^{-4}\,\mathrm{yr}^{-1}$ at the present day. For comparison, \citet{maoz2010} reported a combined LMC and SMC rate of $(2.5$--$4.6)\times10^{-3}\,\mathrm{yr}^{-1}$. The predicted present-day rate is approximately a factor of five below the lower boundary of this range, while the peak model rate is within a factor of two. Because the observational estimate includes both Magellanic Clouds, it is not a direct reference for an LMC-only model. The model prediction is therefore below, but within an order of magnitude of, the available combined estimate.

\begin{figure}
\centering
\includegraphics[width=\columnwidth]{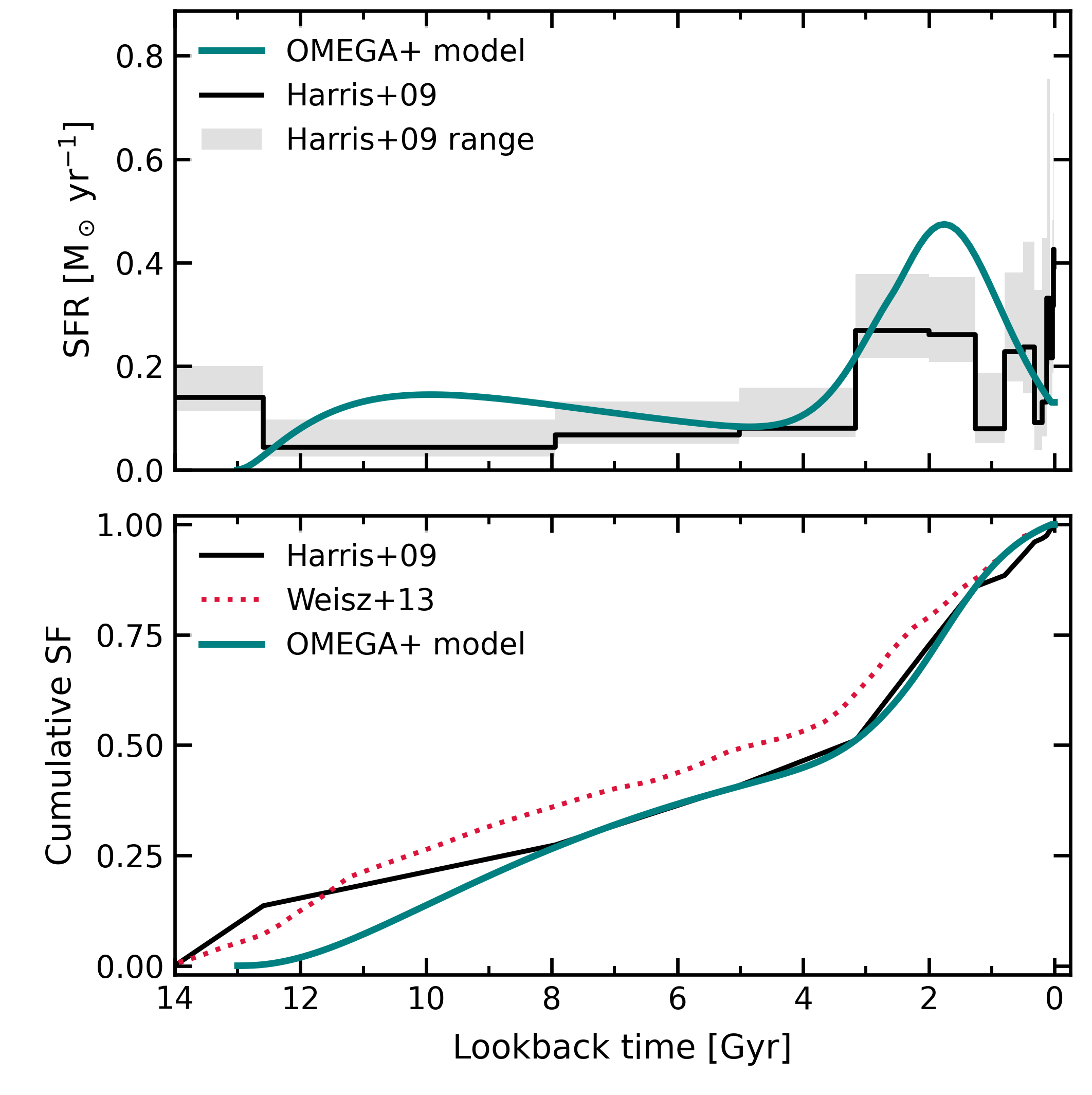}
\caption{Comparison between the selected \textsc{OMEGA+} model and reconstructed LMC SFHs. Top: model SFR and the global \citet{harris2009} bins with their lower and upper limits. Bottom: cumulative star formation compared with \citet{harris2009} and \citet{weisz2013}.}
\label{fig:fig5}
\end{figure}

Figure~\ref{fig:fig5} compares the model directly with the observationally resolved reconstruction of the global SFH by \citet{harris2009}. The model captures the broad late rise and reaches a similar recent SFR, although it does not reproduce every time bin. The main difference occurs at the oldest ages. The Harris reconstruction assigns a larger fraction of stellar mass to the early period, followed by a quieter intermediate interval. The cumulative history from \citet{weisz2013} similarly rises earlier than the model. These differences show that the abundance track does not uniquely determine the time distribution of star formation. The delayed modelled buildup may also contribute to the MDF discrepancy discussed in Sect.~\ref{subsec:chem}, because the model continues to form stars while its metallicity is still low.

The posterior diagnostics provide several encouraging consistency checks. The model reproduces the Mg and O sequences, places the principal Mg, O, and MDF peaks close to their observed positions, gives a present SFR within the adopted comparison range, and broadly reproduces the timing of the late SFH enhancement. The present-day stellar mass is also in excellent agreement with observation-based estimates. Despite these agreements, there are also some model predictions that do not fully match the observations. The model MDF is broader than the observed one, especially in the metal-poor region, and the model also predicts higher mass for the gas content of LMC than what is observed.

\subsection{Comparison with literature results}

The selected history combines two mechanisms previously discussed separately for the LMC. \citet{nidever2020} and \citet{hasselquist2021} generated the APOGEE Mg upturn with explicit SFE enhancements, while models of the LMC--SMC interaction allow late transfer of metal-poor or enriched gas \citep{bekki2007,bekki2012}. In our fit, the second infall increases the fuel supply before the SFE enhancement raises the CCSN contribution.

The event timing is consistent with earlier abundance-based studies. The Gaussian SFE enhancement in our model is centred at $t=11.91$~Gyr and the SFR peaks at 11.25~Gyr. The model of \citet{nidever2020} reaches its SFR maximum at 12.25~Gyr. In the \textsc{flexCE} LMC model of \citet{hasselquist2021}, the fitted SFE enhancement is centred at 11~Gyr and the SFR reaches its late maximum at approximately 12~Gyr. The parameterisations are different, so the SFE amplitudes cannot be compared directly. The late accreted mass in our normalised model is approximately $2\times10^9\,\Msun$, which is of the same order as the external gas masses considered by \citet{bekki2012}. However, the authors noted that existing tidal simulations transferred less than $10^8\,\Msun$ from an SMC with approximately its present-day mass, substantially below the mass required to produce the proposed metallicity decrease in the age–metallicity relation. Our inferred late-infall mass is therefore larger than expected from those simulations. The result supports late accretion of gas with an SMC-like composition, but it does not demonstrate that the entire inferred mass was transferred from the SMC. Such an event would require a more massive or gas-rich past SMC, another external gas reservoir, or a combination of these sources.

The outflow loading also lies within the broad range explored by earlier abundance-based models. The selected value, $\eta=5.184$, falls between $\eta=2$ in the preferred \textsc{flexCE} solution of \citet{hasselquist2021} and $\eta=10$ in the model of \citet{nidever2020}. This spread shows that the Mg-track morphology alone does not determine the outflow efficiency tightly. Changes in $\eta$ can be compensated by changes in gas supply, SFE, IMF, SN~Ia enrichment, and yields. Our value should consequently be interpreted as part of the selected parameter combination and not as an independent measurement of the LMC wind loading.

The MDF remains an important discriminator. \citet{nidever2020} and \citet{hasselquist2021} fitted the abundance tracks and subsequently constructed the MDFs as posterior comparisons. Both studies discuss the effect of APOGEE selection. In additional control calculations, we retained the second-infall history and the other parameters of the hybrid models but removed the temporary SFE enhancement. These otherwise identical late-infall-only models generally diluted the abundance track instead of producing a strong Mg rise. Making the late event more massive or more strongly dominated by primordial gas produced deeper dilution and could generate loops or extended metal-rich tails as subsequent star formation enriched the gas again. The selected hybrid model avoids those failures, but it still forms too many metal-poor stars and produces an MDF that is too broad. This shows that improving the median track and SFH does not guarantee a correct stellar distribution.

The present-day mass components provide another independent test. Our model gives $M_\star=1.88\times10^9\,\Msun$, in excellent agreement with observation-based estimates of approximately $(1.9$--$2.0)\times10^9\,\Msun$ and within the wider published range of $(1.5$--$2.7)\times10^9\,\Msun$. The preferred \textsc{flexCE} solution of \citet{hasselquist2021} yielded approximately $7.3\times10^9\,\Msun$. However, its absolute mass scale was not adjusted to reproduce the observed final stellar mass, so this is not a one-to-one comparison with our model prediction. Our model retains more gas than indicated by the adopted neutral-gas estimate, making the gas fraction a useful target for future refinement.

The fitted parameters are not a unique reconstruction of LMC history. Several combinations of gas supply, SFE, outflow loading, IMF, SN~Ia production, and yields can produce similar abundance tracks. The result provides evidence that mixed late gas accretion combined with enhanced star formation is a plausible channel for the Mg feature. It does not prove that this exact parameter vector describes the LMC. The remaining MDF, mass-partition, and SN-rate discrepancies reinforce this interpretation.

\section{Conclusions}
\label{sec:conclusions}

We modelled the APOGEE-2 DR17 LMC RGB abundance sequence with a hybrid two-infall \textsc{OMEGA+} model. The formal fit used only the Mg median sequence. All population and time-domain quantities were evaluated afterwards. Our main conclusions are as follows:

\begin{enumerate}
    \item The hybrid model reproduces the Mg sequence with an RMS residual of 0.010~dex. It also reproduces O closely and gives an acceptable Ca sequence, while Si remains less successful.

    \item The best-fitting history contains a second, mixed-gas infall beginning at $t=10.391$~Gyr with $M_2/M_1=0.159$, followed by a Gaussian SFE enhancement centred at 11.906~Gyr. Their combined action produces the observed late Mg upturn. The result demonstrates that a two-infall interpretation with variable star-formation efficiency is viable.

    \item The model captures the broad timing of the late resolved SFH and gives a present-day SFR of $0.124\,\Msunyr$, within the adopted observational comparison range. The observational cumulative SFHs place a larger fraction of star formation at early times, but the model reproduces the principal late enhancement associated with the Mg upturn.

    \item The present-day stellar mass agrees with observation-based estimates, while the model also gives a present SFR consistent with the observations. The model retains more gas than indicated by the neutral-gas estimate, and its predicted SN rate lies below the available combined LMC--SMC estimate.

    \item The model reproduces the position of the MDF peak and places the Mg and O distribution peaks close to their observed positions. However, the model MDF is broader and more metal-poor than the APOGEE RGB distribution, with an excess of stars at approximately $-2\lesssim\feh\lesssim-1$. Because no APOGEE RGB selection function was applied to the model, these distributions remain posterior qualitative diagnostics. A future population-level analysis should include the relative number of stars formed at different metallicities together with a forward model of the observational selection.
\end{enumerate}

Our model therefore provides a physically plausible hybrid route to the LMC Mg feature, but it still predicts too much gas content. A future population-level fit should forward-model the RGB selection function and fit the abundance track, MDF, and other stellar distributions consistently.

\begin{acknowledgements}

Funding for the Sloan Digital Sky Survey IV has been provided by the Alfred P. Sloan Foundation, the U.S. Department of Energy Office of Science, and the Participating Institutions.

SDSS-IV acknowledges support and resources from the Center for High Performance Computing at the University of Utah. The SDSS website is www.sdss.org.

SDSS-IV is managed by the Astrophysical Research Consortium for the Participating Institutions of the SDSS Collaboration including the Brazilian Participation Group, the Carnegie Institution for Science, Carnegie Mellon University, Center for Astrophysics | Harvard \& Smithsonian, the Chilean Participation Group, the French Participation Group, Instituto de Astrof\'isica de Canarias, The Johns Hopkins University, Kavli Institute for the Physics and Mathematics of the Universe (IPMU) / University of Tokyo, the Korean Participation Group, Lawrence Berkeley National Laboratory, Leibniz Institut f\"ur Astrophysik Potsdam (AIP), Max-Planck-Institut f\"ur Astronomie (MPIA Heidelberg), Max-Planck-Institut f\"ur Astrophysik (MPA Garching), Max-Planck-Institut f\"ur Extraterrestrische Physik (MPE), National Astronomical Observatories of China, New Mexico State University, New York University, University of Notre Dame, Observat\'ario Nacional / MCTI, The Ohio State University, Pennsylvania State University, Shanghai Astronomical Observatory, United Kingdom Participation Group, Universidad Nacional Aut\'onoma de M\'exico, University of Arizona, University of Colorado Boulder, University of Oxford, University of Portsmouth, University of Utah, University of Virginia, University of Washington, University of Wisconsin, Vanderbilt University, and Yale University.

\end{acknowledgements}

\bibliographystyle{aa}
\bibliography{references}

@ARTICLE{esienstein2011,
       author = {{Eisenstein}, Daniel J. and {Weinberg}, David H. and {Agol}, Eric and {Aihara}, Hiroaki and {Allende Prieto}, Carlos and {Anderson}, Scott F. and {Arns}, James A. and {Aubourg}, {\'E}ric and {Bailey}, Stephen and {Balbinot}, Eduardo and {Barkhouser}, Robert and {Beers}, Timothy C. and {Berlind}, Andreas A. and {Bickerton}, Steven J. and {Bizyaev}, Dmitry and {Blanton}, Michael R. and {Bochanski}, John J. and {Bolton}, Adam S. and {Bosman}, Casey T. and {Bovy}, Jo and {Brandt}, W.~N. and {Breslauer}, Ben and {Brewington}, Howard J. and {Brinkmann}, J. and {Brown}, Peter J. and {Brownstein}, Joel R. and {Burger}, Dan and {Busca}, Nicolas G. and {Campbell}, Heather and {Cargile}, Phillip A. and {Carithers}, William C. and {Carlberg}, Joleen K. and {Carr}, Michael A. and {Chang}, Liang and {Chen}, Yanmei and {Chiappini}, Cristina and {Comparat}, Johan and {Connolly}, Natalia and {Cortes}, Marina and {Croft}, Rupert A.~C. and {Cunha}, Katia and {da Costa}, Luiz N. and {Davenport}, James R.~A. and {Dawson}, Kyle and {De Lee}, Nathan and {Porto de Mello}, Gustavo F. and {de Simoni}, Fernando and {Dean}, Janice and {Dhital}, Saurav and {Ealet}, Anne and {Ebelke}, Garrett L. and {Edmondson}, Edward M. and {Eiting}, Jacob M. and {Escoffier}, Stephanie and {Esposito}, Massimiliano and {Evans}, Michael L. and {Fan}, Xiaohui and {Femen{\'\i}a Castell{\'a}}, Bruno and {Dutra Ferreira}, Leticia and {Fitzgerald}, Greg and {Fleming}, Scott W. and {Font-Ribera}, Andreu and {Ford}, Eric B. and {Frinchaboy}, Peter M. and {Garc{\'\i}a P{\'e}rez}, Ana Elia and {Gaudi}, B. Scott and {Ge}, Jian and {Ghezzi}, Luan and {Gillespie}, Bruce A. and {Gilmore}, G. and {Girardi}, L{\'e}o and {Gott}, J. Richard and {Gould}, Andrew and {Grebel}, Eva K. and {Gunn}, James E. and {Hamilton}, Jean-Christophe and {Harding}, Paul and {Harris}, David W. and {Hawley}, Suzanne L. and {Hearty}, Frederick R. and {Hennawi}, Joseph F. and {Gonz{\'a}lez Hern{\'a}ndez}, Jonay I. and {Ho}, Shirley and {Hogg}, David W. and {Holtzman}, Jon A. and {Honscheid}, Klaus and {Inada}, Naohisa and {Ivans}, Inese I. and {Jiang}, Linhua and {Jiang}, Peng and {Johnson}, Jennifer A. and {Jordan}, Cathy and {Jordan}, Wendell P. and {Kauffmann}, Guinevere and {Kazin}, Eyal and {Kirkby}, David and {Klaene}, Mark A. and {Knapp}, G.~R. and {Kneib}, Jean-Paul and {Kochanek}, C.~S. and {Koesterke}, Lars and {Kollmeier}, Juna A. and {Kron}, Richard G. and {Lampeitl}, Hubert and {Lang}, Dustin and {Lawler}, James E. and {Le Goff}, Jean-Marc and {Lee}, Brian L. and {Lee}, Young Sun and {Leisenring}, Jarron M. and {Lin}, Yen-Ting and {Liu}, Jian and {Long}, Daniel C. and {Loomis}, Craig P. and {Lucatello}, Sara and {Lundgren}, Britt and {Lupton}, Robert H. and {Ma}, Bo and {Ma}, Zhibo and {MacDonald}, Nicholas and {Mack}, Claude and {Mahadevan}, Suvrath and {Maia}, Marcio A.~G. and {Majewski}, Steven R. and {Makler}, Martin and {Malanushenko}, Elena and {Malanushenko}, Viktor and {Mandelbaum}, Rachel and {Maraston}, Claudia and {Margala}, Daniel and {Maseman}, Paul and {Masters}, Karen L. and {McBride}, Cameron K. and {McDonald}, Patrick and {McGreer}, Ian D. and {McMahon}, Richard G. and {Mena Requejo}, Olga and {M{\'e}nard}, Brice and {Miralda-Escud{\'e}}, Jordi and {Morrison}, Heather L. and {Mullally}, Fergal and {Muna}, Demitri and {Murayama}, Hitoshi and {Myers}, Adam D. and {Naugle}, Tracy and {Neto}, Angelo Fausti and {Nguyen}, Duy Cuong and {Nichol}, Robert C. and {Nidever}, David L. and {O'Connell}, Robert W. and {Ogando}, Ricardo L.~C. and {Olmstead}, Matthew D. and {Oravetz}, Daniel J. and {Padmanabhan}, Nikhil and {Paegert}, Martin and {Palanque-Delabrouille}, Nathalie and {Pan}, Kaike and {Pandey}, Parul and {Parejko}, John K. and {P{\^a}ris}, Isabelle and {Pellegrini}, Paulo and {Pepper}, Joshua and {Percival}, Will J. and {Petitjean}, Patrick and {Pfaffenberger}, Robert and {Pforr}, Janine and {Phleps}, Stefanie and {Pichon}, Christophe and {Pieri}, Matthew M. and {Prada}, Francisco and {Price-Whelan}, Adrian M. and {Raddick}, M. Jordan and {Ramos}, Beatriz H.~F. and {Reid}, I. Neill and {Reyle}, Celine and {Rich}, James and {Richards}, Gordon T. and {Rieke}, George H. and {Rieke}, Marcia J. and {Rix}, Hans-Walter and {Robin}, Annie C. and {Rocha-Pinto}, Helio J. and {Rockosi}, Constance M. and {Roe}, Natalie A. and {Rollinde}, Emmanuel and {Ross}, Ashley J. and {Ross}, Nicholas P. and {Rossetto}, Bruno and {S{\'a}nchez}, Ariel G. and {Santiago}, Basilio and {Sayres}, Conor and {Schiavon}, Ricardo and {Schlegel}, David J. and {Schlesinger}, Katharine J. and {Schmidt}, Sarah J. and {Schneider}, Donald P. and {Sellgren}, Kris and {Shelden}, Alaina and {Sheldon}, Erin and {Shetrone}, Matthew},
        title = "{SDSS-III: Massive Spectroscopic Surveys of the Distant Universe, the Milky Way, and Extra-Solar Planetary Systems}",
      journal = {AJ},
         year = 2011,
        month = sep,
       volume = {142},
       number = {3},
          eid = {72},
        pages = {72},
          doi = {10.1088/0004-6256/142/3/72},
archivePrefix = {arXiv},
       eprint = {1101.1529},
 primaryClass = {astro-ph.IM},
       adsurl = {https://ui.adsabs.harvard.edu/abs/2011AJ....142...72E}
}

@ARTICLE{blanton2017,
       author = {{Blanton}, Michael R. and {Bershady}, Matthew A. and {Abolfathi}, Bela and {Albareti}, Franco D. and {Allende Prieto}, Carlos and {Almeida}, Andres and {Alonso-Garc{\'\i}a}, Javier and {Anders}, Friedrich and {Anderson}, Scott F. and {Andrews}, Brett and {Aquino-Ort{\'\i}z}, Erik and {Arag{\'o}n-Salamanca}, Alfonso and {Argudo-Fern{\'a}ndez}, Maria and {Armengaud}, Eric and {Aubourg}, Eric and {Avila-Reese}, Vladimir and {Badenes}, Carles and {Bailey}, Stephen and {Barger}, Kathleen A. and {Barrera-Ballesteros}, Jorge and {Bartosz}, Curtis and {Bates}, Dominic and {Baumgarten}, Falk and {Bautista}, Julian and {Beaton}, Rachael and {Beers}, Timothy C. and {Belfiore}, Francesco and {Bender}, Chad F. and {Berlind}, Andreas A. and {Bernardi}, Mariangela and {Beutler}, Florian and {Bird}, Jonathan C. and {Bizyaev}, Dmitry and {Blanc}, Guillermo A. and {Blomqvist}, Michael and {Bolton}, Adam S. and {Boquien}, M{\'e}d{\'e}ric and {Borissova}, Jura and {van den Bosch}, Remco and {Bovy}, Jo and {Brandt}, William N. and {Brinkmann}, Jonathan and {Brownstein}, Joel R. and {Bundy}, Kevin and {Burgasser}, Adam J. and {Burtin}, Etienne and {Busca}, Nicol{\'a}s G. and {Cappellari}, Michele and {Delgado Carigi}, Maria Leticia and {Carlberg}, Joleen K. and {Carnero Rosell}, Aurelio and {Carrera}, Ricardo and {Chanover}, Nancy J. and {Cherinka}, Brian and {Cheung}, Edmond and {G{\'o}mez Maqueo Chew}, Yilen and {Chiappini}, Cristina and {Choi}, Peter Doohyun and {Chojnowski}, Drew and {Chuang}, Chia-Hsun and {Chung}, Haeun and {Cirolini}, Rafael Fernando and {Clerc}, Nicolas and {Cohen}, Roger E. and {Comparat}, Johan and {da Costa}, Luiz and {Cousinou}, Marie-Claude and {Covey}, Kevin and {Crane}, Jeffrey D. and {Croft}, Rupert A.~C. and {Cruz-Gonzalez}, Irene and {Garrido Cuadra}, Daniel and {Cunha}, Katia and {Damke}, Guillermo J. and {Darling}, Jeremy and {Davies}, Roger and {Dawson}, Kyle and {de la Macorra}, Axel and {Dell'Agli}, Flavia and {De Lee}, Nathan and {Delubac}, Timoth{\'e}e and {Di Mille}, Francesco and {Diamond-Stanic}, Aleks and {Cano-D{\'\i}az}, Mariana and {Donor}, John and {Downes}, Juan Jos{\'e} and {Drory}, Niv and {du Mas des Bourboux}, H{\'e}lion and {Duckworth}, Christopher J. and {Dwelly}, Tom and {Dyer}, Jamie and {Ebelke}, Garrett and {Eigenbrot}, Arthur D. and {Eisenstein}, Daniel J. and {Emsellem}, Eric and {Eracleous}, Mike and {Escoffier}, Stephanie and {Evans}, Michael L. and {Fan}, Xiaohui and {Fern{\'a}ndez-Alvar}, Emma and {Fernandez-Trincado}, J.~G. and {Feuillet}, Diane K. and {Finoguenov}, Alexis and {Fleming}, Scott W. and {Font-Ribera}, Andreu and {Fredrickson}, Alexander and {Freischlad}, Gordon and {Frinchaboy}, Peter M. and {Fuentes}, Carla E. and {Galbany}, Llu{\'\i}s and {Garcia-Dias}, R. and {Garc{\'\i}a-Hern{\'a}ndez}, D.~A. and {Gaulme}, Patrick and {Geisler}, Doug and {Gelfand}, Joseph D. and {Gil-Mar{\'\i}n}, H{\'e}ctor and {Gillespie}, Bruce A. and {Goddard}, Daniel and {Gonzalez-Perez}, Violeta and {Grabowski}, Kathleen and {Green}, Paul J. and {Grier}, Catherine J. and {Gunn}, James E. and {Guo}, Hong and {Guy}, Julien and {Hagen}, Alex and {Hahn}, ChangHoon and {Hall}, Matthew and {Harding}, Paul and {Hasselquist}, Sten and {Hawley}, Suzanne L. and {Hearty}, Fred and {Gonzalez Hern{\'a}ndez}, Jonay I. and {Ho}, Shirley and {Hogg}, David W. and {Holley-Bockelmann}, Kelly and {Holtzman}, Jon A. and {Holzer}, Parker H. and {Huehnerhoff}, Joseph and {Hutchinson}, Timothy A. and {Hwang}, Ho Seong and {Ibarra-Medel}, H{\'e}ctor J. and {da Silva Ilha}, Gabriele and {Ivans}, Inese I. and {Ivory}, KeShawn and {Jackson}, Kelly and {Jensen}, Trey W. and {Johnson}, Jennifer A. and {Jones}, Amy and {J{\"o}nsson}, Henrik and {Jullo}, Eric and {Kamble}, Vikrant and {Kinemuchi}, Karen and {Kirkby}, David and {Kitaura}, Francisco-Shu and {Klaene}, Mark and {Knapp}, Gillian R. and {Kneib}, Jean-Paul and {Kollmeier}, Juna A. and {Lacerna}, Ivan and {Lane}, Richard R. and {Lang}, Dustin and {Law}, David R. and {Lazarz}, Daniel and {Lee}, Youngbae and {Le Goff}, Jean-Marc and {Liang}, Fu-Heng and {Li}, Cheng and {Li}, Hongyu and {Lian}, Jianhui and {Lima}, Marcos and {Lin}, Lihwai and {Lin}, Yen-Ting and {Bertran de Lis}, Sara and {Liu}, Chao and {de Icaza Lizaola}, Miguel Angel C. and {Long}, Dan and {Lucatello}, Sara and {Lundgren}, Britt and {MacDonald}, Nicholas K. and {Deconto Machado}, Alice and {MacLeod}, Chelsea L. and {Mahadevan}, Suvrath and {Geimba Maia}, Marcio Antonio and {Maiolino}, Roberto and {Majewski}, Steven R. and {Malanushenko}, Elena and {Malanushenko}, Viktor and {Manchado}, Arturo and {Mao}, Shude and {Maraston}, Claudia and {Marques-Chaves}, Rui and {Masseron}, Thomas and {Masters}, Karen L. and {McBride}, Cameron K. and {McDermid}, Richard M. and {McGrath}, Brianne and {McGreer}, Ian D. and {Medina Pe{\~n}a}, Nicol{\'a}s and {Melendez}, Matthew and {Merloni}, Andrea and {Merrifield}, Michael R. and {Meszaros}, Szabolcs and {Meza}, Andres and {Minchev}, Ivan and {Minniti}, Dante and {Miyaji}, Takamitsu and {More}, Surhud and {Mulchaey}, John and {M{\"u}ller-S{\'a}nchez}, Francisco and {Muna}, Demitri and {Munoz}, Ricardo R. and {Myers}, Adam D. and {Nair}, Preethi and {Nandra}, Kirpal and {Correa do Nascimento}, Janaina and {Negrete}, Alenka and {Ness}, Melissa and {Newman}, Jeffrey A. and {Nichol}, Robert C. and {Nidever}, David L. and {Nitschelm}, Christian and {Ntelis}, Pierros and {O'Connell}, Julia E. and {Oelkers}, Ryan J. and {Oravetz}, Audrey and {Oravetz}, Daniel and {Pace}, Zach and {Padilla}, Nelson and {Palanque-Delabrouille}, Nathalie and {Alonso Palicio}, Pedro and {Pan}, Kaike and {Parejko}, John K. and {Parikh}, Taniya and {P{\^a}ris}, Isabelle and {Park}, Changbom and {Patten}, Alim Y. and {Peirani}, Sebastien and {Pellejero-Ibanez}, Marcos and {Penny}, Samantha and {Percival}, Will J. and {Perez-Fournon}, Ismael and {Petitjean}, Patrick and {Pieri}, Matthew M. and {Pinsonneault}, Marc and {Pisani}, Alice and {Poleski}, Rados{\l}aw and {Prada}, Francisco and {Prakash}, Abhishek and {Queiroz}, Anna B{\'a}rbara de Andrade and {Raddick}, M. Jordan and {Raichoor}, Anand and {Barboza Rembold}, Sandro and {Richstein}, Hannah and {Riffel}, Rogemar A. and {Riffel}, Rog{\'e}rio and {Rix}, Hans-Walter and {Robin}, Annie C. and {Rockosi}, Constance M. and {Rodr{\'\i}guez-Torres}, Sergio and {Roman-Lopes}, A. and {Rom{\'a}n-Z{\'u}{\~n}iga}, Carlos and {Rosado}, Margarita and {Ross}, Ashley J. and {Rossi}, Graziano and {Ruan}, John and {Ruggeri}, Rossana and {Rykoff}, Eli S. and {Salazar-Albornoz}, Salvador and {Salvato}, Mara and {S{\'a}nchez}, Ariel G. and {Aguado}, D.~S. and {S{\'a}nchez-Gallego}, Jos{\'e} R. and {Santana}, Felipe A. and {Santiago}, Bas{\'\i}lio Xavier and {Sayres}, Conor and {Schiavon}, Ricardo P. and {da Silva Schimoia}, Jaderson and {Schlafly}, Edward F. and {Schlegel}, David J. and {Schneider}, Donald P. and {Schultheis}, Mathias and {Schuster}, William J. and {Schwope}, Axel and {Seo}, Hee-Jong and {Shao}, Zhengyi and {Shen}, Shiyin and {Shetrone}, Matthew and {Shull}, Michael and {Simon}, Joshua D. and {Skinner}, Danielle and {Skrutskie}, M.~F. and {Slosar}, An{\v{z}}e and {Smith}, Verne V. and {Sobeck}, Jennifer S. and {Sobreira}, Flavia and {Somers}, Garrett and {Souto}, Diogo and {Stark}, David V. and {Stassun}, Keivan and {Stauffer}, Fritz and {Steinmetz}, Matthias and {Storchi-Bergmann}, Thaisa and {Streblyanska}, Alina and {Stringfellow}, Guy S. and {Su{\'a}rez}, Genaro and {Sun}, Jing and {Suzuki}, Nao and {Szigeti}, Laszlo and {Taghizadeh-Popp}, Manuchehr and {Tang}, Baitian and {Tao}, Charling and {Tayar}, Jamie and {Tembe}, Mita and {Teske}, Johanna and {Thakar}, Aniruddha R. and {Thomas}, Daniel and {Thompson}, Benjamin A. and {Tinker}, Jeremy L. and {Tissera}, Patricia and {Tojeiro}, Rita and {Hernandez Toledo}, Hector and {de la Torre}, Sylvain and {Tremonti}, Christy and {Troup}, Nicholas W. and {Valenzuela}, Octavio and {Martinez Valpuesta}, Inma and {Vargas-Gonz{\'a}lez}, Jaime and {Vargas-Maga{\~n}a}, Mariana and {Vazquez}, Jose Alberto and {Villanova}, Sandro and {Vivek}, M. and {Vogt}, Nicole and {Wake}, David and {Walterbos}, Rene and {Wang}, Yuting and {Weaver}, Benjamin Alan and {Weijmans}, Anne-Marie and {Weinberg}, David H. and {Westfall}, Kyle B. and {Whelan}, David G. and {Wild}, Vivienne and {Wilson}, John and {Wood-Vasey}, W.~M. and {Wylezalek}, Dominika and {Xiao}, Ting and {Yan}, Renbin and {Yang}, Meng and {Ybarra}, Jason E. and {Y{\`e}che}, Christophe and {Zakamska}, Nadia and {Zamora}, Olga and {Zarrouk}, Pauline and {Zasowski}, Gail and {Zhang}, Kai and {Zhao}, Gong-Bo and {Zheng}, Zheng and {Zheng}, Zheng and {Zhou}, Xu and {Zhou}, Zhi-Min and {Zhu}, Guangtun B. and {Zoccali}, Manuela and {Zou}, Hu},
        title = "{Sloan Digital Sky Survey IV: Mapping the Milky Way, Nearby Galaxies, and the Distant Universe}",
      journal = {AJ},
         year = 2017,
        month = jul,
       volume = {154},
       number = {1},
          eid = {28},
        pages = {28},
          doi = {10.3847/1538-3881/aa7567},
archivePrefix = {arXiv},
       eprint = {1703.00052},
 primaryClass = {astro-ph.GA},
       adsurl = {https://ui.adsabs.harvard.edu/abs/2017AJ....154...28B}
}

@ARTICLE{majewski2017,
       author = {{Majewski}, Steven R. and {Schiavon}, Ricardo P. and {Frinchaboy}, Peter M. and {Allende Prieto}, Carlos and {Barkhouser}, Robert and {Bizyaev}, Dmitry and {Blank}, Basil and {Brunner}, Sophia and {Burton}, Adam and {Carrera}, Ricardo and {Chojnowski}, S. Drew and {Cunha}, K{\'a}tia and {Epstein}, Courtney and {Fitzgerald}, Greg and {Garc{\'\i}a P{\'e}rez}, Ana E. and {Hearty}, Fred R. and {Henderson}, Chuck and {Holtzman}, Jon A. and {Johnson}, Jennifer A. and {Lam}, Charles R. and {Lawler}, James E. and {Maseman}, Paul and {M{\'e}sz{\'a}ros}, Szabolcs and {Nelson}, Matthew and {Nguyen}, Duy Coung and {Nidever}, David L. and {Pinsonneault}, Marc and {Shetrone}, Matthew and {Smee}, Stephen and {Smith}, Verne V. and {Stolberg}, Todd and {Skrutskie}, Michael F. and {Walker}, Eric and {Wilson}, John C. and {Zasowski}, Gail and {Anders}, Friedrich and {Basu}, Sarbani and {Beland}, Stephane and {Blanton}, Michael R. and {Bovy}, Jo and {Brownstein}, Joel R. and {Carlberg}, Joleen and {Chaplin}, William and {Chiappini}, Cristina and {Eisenstein}, Daniel J. and {Elsworth}, Yvonne and {Feuillet}, Diane and {Fleming}, Scott W. and {Galbraith-Frew}, Jessica and {Garc{\'\i}a}, Rafael A. and {Garc{\'\i}a-Hern{\'a}ndez}, D. An{\'\i}bal and {Gillespie}, Bruce A. and {Girardi}, L{\'e}o and {Gunn}, James E. and {Hasselquist}, Sten and {Hayden}, Michael R. and {Hekker}, Saskia and {Ivans}, Inese and {Kinemuchi}, Karen and {Klaene}, Mark and {Mahadevan}, Suvrath and {Mathur}, Savita and {Mosser}, Beno{\^\i}t and {Muna}, Demitri and {Munn}, Jeffrey A. and {Nichol}, Robert C. and {O'Connell}, Robert W. and {Parejko}, John K. and {Robin}, A.~C. and {Rocha-Pinto}, Helio and {Schultheis}, Matthias and {Serenelli}, Aldo M. and {Shane}, Neville and {Silva Aguirre}, Victor and {Sobeck}, Jennifer S. and {Thompson}, Benjamin and {Troup}, Nicholas W. and {Weinberg}, David H. and {Zamora}, Olga},
        title = "{The Apache Point Observatory Galactic Evolution Experiment (APOGEE)}",
      journal = {AJ},
         year = 2017,
        month = sep,
       volume = {154},
       number = {3},
          eid = {94},
        pages = {94},
          doi = {10.3847/1538-3881/aa784d},
archivePrefix = {arXiv},
       eprint = {1509.05420},
 primaryClass = {astro-ph.IM},
       adsurl = {https://ui.adsabs.harvard.edu/abs/2017AJ....154...94M}
}

@article{garciaperez2016,
doi = {10.3847/0004-6256/151/6/144},
url = {https://dx.doi.org/10.3847/0004-6256/151/6/144},
year = {2016},
month = {may},
publisher = {The American Astronomical Society},
volume = {151},
number = {6},
pages = {144},
author = {García Pérez, Ana E. and Prieto, Carlos Allende and Holtzman, Jon A. and Shetrone, Matthew and Mészáros, Szabolcs and Bizyaev, Dmitry and Carrera, Ricardo and Cunha, Katia and García-Hernández, D. A. and Johnson, Jennifer A. and Majewski, Steven R. and Nidever, David L. and Schiavon, Ricardo P. and Shane, Neville and Smith, Verne V. and Sobeck, Jennifer and Troup, Nicholas and Zamora, Olga and Weinberg, David H. and Bovy, Jo and Eisenstein, Daniel J. and Feuillet, Diane and Frinchaboy, Peter M. and Hayden, Michael R. and Hearty, Fred R. and Nguyen, Duy C. and O’Connell, Robert W. and Pinsonneault, Marc H. and Wilson, John C. and Zasowski, Gail},
title = {ASPCAP: THE APOGEE STELLAR PARAMETER AND CHEMICAL ABUNDANCES PIPELINE},
journal = {AJ}
}

@ARTICLE{abdurrouf2022,
       author = {{Abdurro'uf} and {Accetta}, Katherine and {Aerts}, Conny and {Silva Aguirre}, V{\'\i}ctor and {Ahumada}, Romina and {Ajgaonkar}, Nikhil and {Filiz Ak}, N. and {Alam}, Shadab and {Allende Prieto}, Carlos and {Almeida}, Andr{\'e}s and {Anders}, Friedrich and {Anderson}, Scott F. and {Andrews}, Brett H. and {Anguiano}, Borja and {Aquino-Ort{\'\i}z}, Erik and {Arag{\'o}n-Salamanca}, Alfonso and {Argudo-Fern{\'a}ndez}, Maria and {Ata}, Metin and {Aubert}, Marie and {Avila-Reese}, Vladimir and {Badenes}, Carles and {Barb{\'a}}, Rodolfo H. and {Barger}, Kat and {Barrera-Ballesteros}, Jorge K. and {Beaton}, Rachael L. and {Beers}, Timothy C. and {Belfiore}, Francesco and {Bender}, Chad F. and {Bernardi}, Mariangela and {Bershady}, Matthew A. and {Beutler}, Florian and {Bidin}, Christian Moni and {Bird}, Jonathan C. and {Bizyaev}, Dmitry and {Blanc}, Guillermo A. and {Blanton}, Michael R. and {Boardman}, Nicholas Fraser and {Bolton}, Adam S. and {Boquien}, M{\'e}d{\'e}ric and {Borissova}, Jura and {Bovy}, Jo and {Brandt}, W.~N. and {Brown}, Jordan and {Brownstein}, Joel R. and {Brusa}, Marcella and {Buchner}, Johannes and {Bundy}, Kevin and {Burchett}, Joseph N. and {Bureau}, Martin and {Burgasser}, Adam and {Cabang}, Tuesday K. and {Campbell}, Stephanie and {Cappellari}, Michele and {Carlberg}, Joleen K. and {Wanderley}, F{\'a}bio Carneiro and {Carrera}, Ricardo and {Cash}, Jennifer and {Chen}, Yan-Ping and {Chen}, Wei-Huai and {Cherinka}, Brian and {Chiappini}, Cristina and {Choi}, Peter Doohyun and {Chojnowski}, S. Drew and {Chung}, Haeun and {Clerc}, Nicolas and {Cohen}, Roger E. and {Comerford}, Julia M. and {Comparat}, Johan and {da Costa}, Luiz and {Covey}, Kevin and {Crane}, Jeffrey D. and {Cruz-Gonzalez}, Irene and {Culhane}, Connor and {Cunha}, Katia and {Dai}, Y. Sophia and {Damke}, Guillermo and {Darling}, Jeremy and {Davidson}, James W., Jr. and {Davies}, Roger and {Dawson}, Kyle and {De Lee}, Nathan and {Diamond-Stanic}, Aleksandar M. and {Cano-D{\'\i}az}, Mariana and {S{\'a}nchez}, Helena Dom{\'\i}nguez and {Donor}, John and {Duckworth}, Chris and {Dwelly}, Tom and {Eisenstein}, Daniel J. and {Elsworth}, Yvonne P. and {Emsellem}, Eric and {Eracleous}, Mike and {Escoffier}, Stephanie and {Fan}, Xiaohui and {Farr}, Emily and {Feng}, Shuai and {Fern{\'a}ndez-Trincado}, Jos{\'e} G. and {Feuillet}, Diane and {Filipp}, Andreas and {Fillingham}, Sean P. and {Frinchaboy}, Peter M. and {Fromenteau}, Sebastien and {Galbany}, Llu{\'\i}s and {Garc{\'\i}a}, Rafael A. and {Garc{\'\i}a-Hern{\'a}ndez}, D.~A. and {Ge}, Junqiang and {Geisler}, Doug and {Gelfand}, Joseph and {G{\'e}ron}, Tobias and {Gibson}, Benjamin J. and {Goddy}, Julian and {Godoy-Rivera}, Diego and {Grabowski}, Kathleen and {Green}, Paul J. and {Greener}, Michael and {Grier}, Catherine J. and {Griffith}, Emily and {Guo}, Hong and {Guy}, Julien and {Hadjara}, Massinissa and {Harding}, Paul and {Hasselquist}, Sten and {Hayes}, Christian R. and {Hearty}, Fred and {Hern{\'a}ndez}, Jes{\'u}s and {Hill}, Lewis and {Hogg}, David W. and {Holtzman}, Jon A. and {Horta}, Danny and {Hsieh}, Bau-Ching and {Hsu}, Chin-Hao and {Hsu}, Yun-Hsin and {Huber}, Daniel and {Huertas-Company}, Marc and {Hutchinson}, Brian and {Hwang}, Ho Seong and {Ibarra-Medel}, H{\'e}ctor J. and {Chitham}, Jacob Ider and {Ilha}, Gabriele S. and {Imig}, Julie and {Jaekle}, Will and {Jayasinghe}, Tharindu and {Ji}, Xihan and {Johnson}, Jennifer A. and {Jones}, Amy and {J{\"o}nsson}, Henrik and {Katkov}, Ivan and {Khalatyan}, Arman, Dr. and {Kinemuchi}, Karen and {Kisku}, Shobhit and {Knapen}, Johan H. and {Kneib}, Jean-Paul and {Kollmeier}, Juna A. and {Kong}, Miranda and {Kounkel}, Marina and {Kreckel}, Kathryn and {Krishnarao}, Dhanesh and {Lacerna}, Ivan and {Lane}, Richard R. and {Langgin}, Rachel and {Lavender}, Ramon and {Law}, David R. and {Lazarz}, Daniel and {Leung}, Henry W. and {Leung}, Ho-Hin and {Lewis}, Hannah M. and {Li}, Cheng and {Li}, Ran and {Lian}, Jianhui and {Liang}, Fu-Heng and {Lin}, Lihwai and {Lin}, Yen-Ting and {Lin}, Sicheng and {Lintott}, Chris and {Long}, Dan and {Longa-Pe{\~n}a}, Pen{\'e}lope and {L{\'o}pez-Cob{\'a}}, Carlos and {Lu}, Shengdong and {Lundgren}, Britt F. and {Luo}, Yuanze and {Mackereth}, J. Ted and {de la Macorra}, Axel and {Mahadevan}, Suvrath and {Majewski}, Steven R. and {Manchado}, Arturo and {Mandeville}, Travis and {Maraston}, Claudia and {Margalef-Bentabol}, Berta and {Masseron}, Thomas and {Masters}, Karen L. and {Mathur}, Savita and {McDermid}, Richard M. and {Mckay}, Myles and {Merloni}, Andrea and {Merrifield}, Michael and {Meszaros}, Szabolcs and {Miglio}, Andrea and {Di Mille}, Francesco and {Minniti}, Dante and {Minsley}, Rebecca and {Monachesi}, Antonela and {Moon}, Jeongin and {Mosser}, Benoit and {Mulchaey}, John and {Muna}, Demitri and {Mu{\~n}oz}, Ricardo R. and {Myers}, Adam D. and {Myers}, Natalie and {Nadathur}, Seshadri and {Nair}, Preethi and {Nandra}, Kirpal and {Neumann}, Justus and {Newman}, Jeffrey A. and {Nidever}, David L. and {Nikakhtar}, Farnik and {Nitschelm}, Christian and {O'Connell}, Julia E. and {Garma-Oehmichen}, Luis and {Luan Souza de Oliveira}, Gabriel and {Olney}, Richard and {Oravetz}, Daniel and {Ortigoza-Urdaneta}, Mario and {Osorio}, Yeisson and {Otter}, Justin and {Pace}, Zachary J. and {Padilla}, Nelson and {Pan}, Kaike and {Pan}, Hsi-An and {Parikh}, Taniya and {Parker}, James and {Peirani}, Sebastien and {Pe{\~n}a Ram{\'\i}rez}, Karla and {Penny}, Samantha and {Percival}, Will J. and {Perez-Fournon}, Ismael and {Pinsonneault}, Marc and {Poidevin}, Fr{\'e}d{\'e}rick and {Poovelil}, Vijith Jacob and {Price-Whelan}, Adrian M. and {B{\'a}rbara de Andrade Queiroz}, Anna and {Raddick}, M. Jordan and {Ray}, Amy and {Rembold}, Sandro Barboza and {Riddle}, Nicole and {Riffel}, Rogemar A. and {Riffel}, Rog{\'e}rio and {Rix}, Hans-Walter and {Robin}, Annie C. and {Rodr{\'\i}guez-Puebla}, Aldo and {Roman-Lopes}, Alexandre and {Rom{\'a}n-Z{\'u}{\~n}iga}, Carlos and {Rose}, Benjamin and {Ross}, Ashley J. and {Rossi}, Graziano and {Rubin}, Kate H.~R. and {Salvato}, Mara and {S{\'a}nchez}, Seb{\'a}stian F. and {S{\'a}nchez-Gallego}, Jos{\'e} R. and {Sanderson}, Robyn and {Santana Rojas}, Felipe Antonio and {Sarceno}, Edgar and {Sarmiento}, Regina and {Sayres}, Conor and {Sazonova}, Elizaveta and {Schaefer}, Adam L. and {Schiavon}, Ricardo and {Schlegel}, David J. and {Schneider}, Donald P. and {Schultheis}, Mathias and {Schwope}, Axel and {Serenelli}, Aldo and {Serna}, Javier and {Shao}, Zhengyi and {Shapiro}, Griffin and {Sharma}, Anubhav and {Shen}, Yue and {Shetrone}, Matthew and {Shu}, Yiping and {Simon}, Joshua D. and {Skrutskie}, M.~F. and {Smethurst}, Rebecca and {Smith}, Verne and {Sobeck}, Jennifer and {Spoo}, Taylor and {Sprague}, Dani and {Stark}, David V. and {Stassun}, Keivan G. and {Steinmetz}, Matthias and {Stello}, Dennis and {Stone-Martinez}, Alexander and {Storchi-Bergmann}, Thaisa and {Stringfellow}, Guy S. and {Stutz}, Amelia and {Su}, Yung-Chau and {Taghizadeh-Popp}, Manuchehr and {Talbot}, Michael S. and {Tayar}, Jamie and {Telles}, Eduardo and {Teske}, Johanna and {Thakar}, Ani and {Theissen}, Christopher and {Tkachenko}, Andrew and {Thomas}, Daniel and {Tojeiro}, Rita and {Hernandez Toledo}, Hector and {Troup}, Nicholas W. and {Trump}, Jonathan R. and {Trussler}, James and {Turner}, Jacqueline and {Tuttle}, Sarah and {Unda-Sanzana}, Eduardo and {V{\'a}zquez-Mata}, Jos{\'e} Antonio and {Valentini}, Marica and {Valenzuela}, Octavio and {Vargas-Gonz{\'a}lez}, Jaime and {Vargas-Maga{\~n}a}, Mariana and {Alfaro}, Pablo Vera and {Villanova}, Sandro and {Vincenzo}, Fiorenzo and {Wake}, David and {Warfield}, Jack T. and {Washington}, Jessica Diane and {Weaver}, Benjamin Alan and {Weijmans}, Anne-Marie and {Weinberg}, David H. and {Weiss}, Achim and {Westfall}, Kyle B. and {Wild}, Vivienne and {Wilde}, Matthew C. and {Wilson}, John C. and {Wilson}, Robert F. and {Wilson}, Mikayla and {Wolf}, Julien and {Wood-Vasey}, W.~M. and {Yan}, Renbin and {Zamora}, Olga and {Zasowski}, Gail and {Zhang}, Kai and {Zhao}, Cheng and {Zheng}, Zheng and {Zheng}, Zheng and {Zhu}, Kai},
        title = "{The Seventeenth Data Release of the Sloan Digital Sky Surveys: Complete Release of MaNGA, MaStar, and APOGEE-2 Data}",
      journal = {ApJS},
         year = 2022,
        month = apr,
       volume = {259},
       number = {2},
          eid = {35},
        pages = {35},
          doi = {10.3847/1538-4365/ac4414},
archivePrefix = {arXiv},
       eprint = {2112.02026},
 primaryClass = {astro-ph.GA},
       adsurl = {https://ui.adsabs.harvard.edu/abs/2022ApJS..259...35A}
}

@ARTICLE{hasselquist2021,
       author = {{Hasselquist}, Sten and {Hayes}, Christian R. and {Lian}, Jianhui and {Weinberg}, David H. and {Zasowski}, Gail and {Horta}, Danny and {Beaton}, Rachael and {Feuillet}, Diane K. and {Garro}, Elisa R. and {Gallart}, Carme and {Smith}, Verne V. and {Holtzman}, Jon A. and {Minniti}, Dante and {Lacerna}, Ivan and {Shetrone}, Matthew and {J{\"o}nsson}, Henrik and {Cioni}, Maria-Rosa L. and {Fillingham}, Sean P. and {Cunha}, Katia and {O'Connell}, Robert and {Fern{\'a}ndez-Trincado}, Jos{\'e} G. and {Mu{\~n}oz}, Ricardo R. and {Schiavon}, Ricardo and {Almeida}, Andres and {Anguiano}, Borja and {Beers}, Timothy C. and {Bizyaev}, Dmitry and {Brownstein}, Joel R. and {Cohen}, Roger E. and {Frinchaboy}, Peter and {Garc{\'\i}a-Hern{\'a}ndez}, D.~A. and {Geisler}, Doug and {Lane}, Richard R. and {Majewski}, Steven R. and {Nidever}, David L. and {Nitschelm}, Christian and {Povick}, Joshua and {Price-Whelan}, Adrian and {Roman-Lopes}, Alexandre and {Rosado}, Margarita and {Sobeck}, Jennifer and {Stringfellow}, Guy and {Valenzuela}, Octavio and {Villanova}, Sandro and {Vincenzo}, Fiorenzo},
        title = "{APOGEE Chemical Abundance Patterns of the Massive Milky Way Satellites}",
      journal = {ApJ},
         year = 2021,
        month = dec,
       volume = {923},
       number = {2},
          eid = {172},
        pages = {172},
          doi = {10.3847/1538-4357/ac25f9},
archivePrefix = {arXiv},
       eprint = {2109.05130},
 primaryClass = {astro-ph.GA},
       adsurl = {https://ui.adsabs.harvard.edu/abs/2021ApJ...923..172H}
}

@ARTICLE{nidever2020,
       author = {{Nidever}, David L. and {Hasselquist}, Sten and {Hayes}, Christian R. and {Hawkins}, Keith and {Povick}, Joshua and {Majewski}, Steven R. and {Smith}, Verne V. and {Anguiano}, Borja and {Stringfellow}, Guy S. and {Sobeck}, Jennifer S. and {Cunha}, Katia and {Beers}, Timothy C. and {Bestenlehner}, Joachim M. and {Cohen}, Roger E. and {Garcia-Hernandez}, D.~A. and {J{\"o}nsson}, Henrik and {Nitschelm}, Christian and {Shetrone}, Matthew and {Lacerna}, Ivan and {Allende Prieto}, Carlos and {Beaton}, Rachael L. and {Dell'Agli}, Flavia and {Fern{\'a}ndez-Trincado}, Jos{\'e} G. and {Feuillet}, Diane and {Gallart}, Carme and {Hearty}, Fred R. and {Holtzman}, Jon and {Manchado}, Arturo and {Mu{\~n}oz}, Ricardo R. and {O'Connell}, Robert and {Rosado}, Margarita},
        title = "{The Lazy Giants: APOGEE Abundances Reveal Low Star Formation Efficiencies in the Magellanic Clouds}",
      journal = {ApJ},
         year = 2020,
        month = jun,
       volume = {895},
       number = {2},
          eid = {88},
        pages = {88},
          doi = {10.3847/1538-4357/ab7305},
archivePrefix = {arXiv},
       eprint = {1901.03448},
 primaryClass = {astro-ph.GA},
       adsurl = {https://ui.adsabs.harvard.edu/abs/2020ApJ...895...88N}
}

@ARTICLE{cote2017,
       author = {{C{\^o}t{\'e}}, Benoit and {O'Shea}, Brian W. and {Ritter}, Christian and {Herwig}, Falk and {Venn}, Kim A.},
        title = "{The Impact of Modeling Assumptions in Galactic Chemical Evolution Models}",
      journal = {ApJ},
         year = {2017},
        month = feb,
       volume = {835},
       number = {2},
          eid = {128},
        pages = {128},
          doi = {10.3847/1538-4357/835/2/128},
archivePrefix = {arXiv},
       eprint = {1604.07824},
 primaryClass = {astro-ph.GA},
       adsurl = {https://ui.adsabs.harvard.edu/abs/2017ApJ...835..128C}
}

@article{cote2018,
doi = {10.3847/1538-4357/aabe8f},
url = {https://doi.org/10.3847/1538-4357/aabe8f},
year = {2018},
month = {may},
publisher = {The American Astronomical Society},
volume = {859},
number = {1},
pages = {67},
author = {Côté, Benoit and Silvia, Devin W. and O’Shea, Brian W. and Smith, Britton and Wise, John H.},
title = {Validating Semi-analytic Models of High-redshift Galaxy Formation Using Radiation Hydrodynamical Simulations},
journal = {ApJ}
}

@ARTICLE{matteucci2021,
       author = {{Matteucci}, Francesca},
        title = "{Modelling the chemical evolution of the Milky Way}",
      journal = {A\&A Rev.},
         year = 2021,
        month = dec,
       volume = {29},
       number = {1},
          eid = {5},
        pages = {5},
          doi = {10.1007/s00159-021-00133-8},
archivePrefix = {arXiv},
       eprint = {2106.13145},
 primaryClass = {astro-ph.GA},
       adsurl = {https://ui.adsabs.harvard.edu/abs/2021Astronomy & AstrophysicsRv..29....5M}
}

@ARTICLE{besla2012,
       author = {{Besla}, Gurtina and {Kallivayalil}, Nitya and {Hernquist}, Lars and {van der Marel}, Roeland P. and {Cox}, T.~J. and {Kere{\v{s}}}, Du{\v{s}}an},
        title = "{The role of dwarf galaxy interactions in shaping the Magellanic System and implications for Magellanic Irregulars}",
      journal = {MNRAS},
         year = 2012,
        month = apr,
       volume = {421},
       number = {3},
        pages = {2109-2138},
          doi = {10.1111/j.1365-2966.2012.20466.x},
archivePrefix = {arXiv},
       eprint = {1201.1299},
 primaryClass = {astro-ph.GA},
       adsurl = {https://ui.adsabs.harvard.edu/abs/2012MNRAS.421.2109B}
}

@article{pagel1998,
    author = {Pagel, B. E. J. and Tautvaišienė, G.},
    title = {Chemical evolution of the Magellanic Clouds: analytical models},
    journal = {MNRAS},
    volume = {299},
    number = {2},
    pages = {535-544},
    year = {1998},
    month = {09},
    issn = {0035-8711},
    doi = {10.1046/j.1365-8711.1998.01792.x},
    url = {https://doi.org/10.1046/j.1365-8711.1998.01792.x},
    eprint = {https://academic.oup.com/mnras/article-pdf/299/2/535/3340375/299-2-535.pdf},
}

@article{bekki2012,
doi = {10.1088/0004-637X/761/2/180},
url = {https://dx.doi.org/10.1088/0004-637X/761/2/180},
year = {2012},
month = {dec},
publisher = {The American Astronomical Society},
volume = {761},
number = {2},
pages = {180},
author = {Bekki, Kenji and Tsujimoto, Takuji},
title = {CHEMICAL EVOLUTION OF THE LARGE MAGELLANIC CLOUD},
journal = {ApJ}
}

@article{harris2009,
doi = {10.1088/0004-6256/138/5/1243},
url = {https://doi.org/10.1088/0004-6256/138/5/1243},
year = {2009},
month = {sep},
publisher = {The American Astronomical Society},
volume = {138},
number = {5},
pages = {1243},
author = {Harris, Jason and Zaritsky, Dennis},
title = {THE STAR FORMATION HISTORY OF THE LARGE MAGELLANIC CLOUD},
journal = {AJ}
}

@article{kim1998,
doi = {10.1086/306030},
url = {https://doi.org/10.1086/306030},
year = {1998},
month = {aug},
publisher = {},
volume = {503},
number = {2},
pages = {674},
author = {Kim, Sungeun and Staveley-Smith, Lister and Dopita, Michael A. and Freeman, Ken C. and Sault, Robert J. and Kesteven, Mike J. and McConnell, David},
title = {An H I Aperture Synthesis Mosaic of the Large Magellanic Cloud},
journal = {ApJ}
}

@article{maoz2010,
    author = {Maoz, Dan and Badenes, Carles},
    title = {The supernova rate and delay time distribution in the Magellanic Clouds},
    journal = {MNRAS},
    volume = {407},
    number = {2},
    pages = {1314-1327},
    year = {2010},
    month = {09},
    issn = {0035-8711},
    doi = {10.1111/j.1365-2966.2010.16988.x},
    url = {https://doi.org/10.1111/j.1365-2966.2010.16988.x},
    eprint = {https://academic.oup.com/mnras/article-pdf/407/2/1314/3914644/mnras0407-1314.pdf},
}

@article{bekki2007,
    author = {Bekki, Kenji and Chiba, Masashi},
    title = {The Magellanic squall: gas replenishment from the Small to the Large Magellanic Cloud},
    journal = {MNRAS},
    volume = {381},
    number = {1},
    pages = {L16-L20},
    year = {2007},
    month = {10},
    issn = {1745-3925},
    doi = {10.1111/j.1745-3933.2007.00357.x},
    url = {https://doi.org/10.1111/j.1745-3933.2007.00357.x},
    eprint = {https://academic.oup.com/mnrasl/article-pdf/381/1/L16/57000731/mnrasl_381_1_l16.pdf},
}

@article{vandermarel2002,
doi = {10.1086/343775},
url = {https://doi.org/10.1086/343775},
year = {2002},
month = {nov},
publisher = {},
volume = {124},
number = {5},
pages = {2639},
author = {van der Marel, Roeland P. and Alves, David R. and Hardy, Eduardo and Suntzeff, Nicholas B.},
title = {New Understanding of Large Magellanic Cloud Structure, Dynamics, and
Orbit from Carbon Star Kinematics},
journal = {AJ}
}

@ARTICLE{salpeter1955b,
       author = {{Salpeter}, Edwin E.},
        title = "{The Luminosity Function and Stellar Evolution.}",
      journal = {ApJ},
         year = 1955,
        month = jan,
       volume = {121},
        pages = {161},
          doi = {10.1086/145971},
       adsurl = {https://ui.adsabs.harvard.edu/abs/1955ApJ...121..161S}
}

@ARTICLE{maoz2012,
       author = {{Maoz}, D. and {Mannucci}, F.},
        title = "{Type-Ia Supernova Rates and the Progenitor Problem: A Review}",
      journal = {PASA},
         year = 2012,
        month = jan,
       volume = {29},
       number = {4},
        pages = {447-465},
          doi = {10.1071/AS11052},
archivePrefix = {arXiv},
       eprint = {1111.4492},
 primaryClass = {astro-ph.CO},
       adsurl = {https://ui.adsabs.harvard.edu/abs/2012PASA...29..447M}
}

@misc{lmfit,
  author       = {Newville, Matthew and
                  Otten, Renee and
                  Nelson, Andrew and
                  Stensitzki, Till and
                  Ingargiola, Antonino and
                  Allan, Daniel and
                  Fox, Austin and
                  Carter, Faustin and
                  Rawlik, Michal},
  title        = {{LMFIT}: Non-Linear Least-Squares Minimization and Curve-Fitting for Python},
  year         = {2025},
  month        = jul,
  howpublished = {Zenodo},
  note         = {Version 1.3.4},
  doi          = {10.5281/zenodo.16175987},
  url          = {https://doi.org/10.5281/zenodo.16175987}
}

@article{francois2004,
	author = {{François, P.} and {Matteucci, F.} and {Cayrel, R.} and {Spite, M.} and {Spite, F.} and {Chiappini, C.}},
	title = {The evolution of the Milky Way from
its earliest phases: Constraints on stellar nucleosynthesis},
	DOI= "10.1051/0004-6361:20034140",
	url= "https://doi.org/10.1051/0004-6361:20034140",
	journal = {A\&A},
	year = 2004,
	volume = 421,
	number = 2,
	pages = "613-621",
}

@article{spitoni2021,
	author = {{Spitoni, E.} and {Verma, K.} and {Silva Aguirre, V.} and {Vincenzo, F.} and {Matteucci, F.} and {Vaičekauskaitė, B.} and {Palla, M.} and {Grisoni, V.} and {Calura, F.}},
	title = {APOGEE DR16: A multi-zone chemical evolution model for the Galactic disc based on MCMC methods},
	DOI= "10.1051/0004-6361/202039864",
	url= "https://doi.org/10.1051/0004-6361/202039864",
	journal = {A\&A},
	year = 2021,
	volume = 647,
	pages = "A73",
}

@article{weisz2013,
    author = {Weisz, Daniel R. and Dolphin, Andrew E. and Skillman, Evan D. and Holtzman, Jon and Dalcanton, Julianne J. and Cole, Andrew A. and Neary, Kyle},
    title = {Comparing the ancient star formation histories of the Magellanic Clouds},
    journal = {MNRAS},
    volume = {431},
    number = {1},
    pages = {364-371},
    year = {2013},
    month = {03},
    issn = {0035-8711},
    doi = {10.1093/mnras/stt165},
    url = {https://doi.org/10.1093/mnras/stt165},
    eprint = {https://academic.oup.com/mnras/article-pdf/431/1/364/18240721/stt165.pdf},
}

@article{meschin2014,
    author = {Meschin, I. and Gallart, C. and Aparicio, A. and Hidalgo, S. L. and Monelli, M. and Stetson, P. B. and Carrera, R.},
    title = {Spatially resolved LMC star formation history – I. Outside in evolution of the outer LMC disc},
    journal = {MNRAS},
    volume = {438},
    number = {2},
    pages = {1067-1080},
    year = {2014},
    month = feb,
    issn = {0035-8711},
    doi = {10.1093/mnras/stt2220},
    url = {https://doi.org/10.1093/mnras/stt2220},
    eprint = {https://academic.oup.com/mnras/article-pdf/438/2/1067/18497677/stt2220.pdf},
}

@article{povick2025,
    author = {Povick, Joshua T and Nidever, David L and Majewski, Steven R and Geisler, Doug and Cioni, Maria-Rosa L and Lu, Yuxi(Lucy) and Muñoz, Ricardo and Stringfellow, Guy S and Almeida, Andrés and Longa-Peña, Penélope and Lane, Richard R and Roman-Lopes, Alexandre},
    title = {Revealing the chemical structure of the Magellanic Clouds with APOGEE. II. Abundance gradients of the Large Magellanic Cloud},
    journal = {MNRAS},
    volume = {544},
    number = {1},
    pages = {457-482},
    year = {2025},
    month = {08},
    issn = {0035-8711},
    doi = {10.1093/mnras/staf1428},
    url = {https://doi.org/10.1093/mnras/staf1428},
    eprint = {https://academic.oup.com/mnras/article-pdf/544/1/457/64149392/staf1428.pdf},
}

@article{vasini2023,
    author = {Vasini, Arianna and Matteucci, F and Spitoni, E and Siegert, Thomas},
    title = {Evolution of radioactive elements in the LMC: predictions for future γ-ray surveys},
    journal = {MNRAS},
    volume = {523},
    number = {1},
    pages = {1153-1162},
    year = {2023},
    month = {05},
    issn = {0035-8711},
    doi = {10.1093/mnras/stad1440},
    url = {https://doi.org/10.1093/mnras/stad1440},
    eprint = {https://academic.oup.com/mnras/article-pdf/523/1/1153/50488490/stad1440.pdf},
}

@inproceedings{grevesse1993,
  author = {{Grevesse}, N. and {Noels}, A.},
  title = {Cosmic abundances of the elements},
  booktitle = {Origin and Evolution of the Elements},
  editor = {{Prantzos}, N. and {Vangioni-Flam}, E. and {Casse}, M.},
  year = {1993},
  pages = {15--25},
  publisher = {Cambridge University Press}
}

@article{grevesse2007,
  author = {{Grevesse}, N. and {Asplund}, M. and {Sauval}, A. J.},
  title = {The solar chemical composition},
  journal = {Space Sci. Rev.},
  year = {2007},
  volume = {130},
  pages = {105--114},
  doi = {10.1007/s11214-007-9173-7}
}

@ARTICLE{mcconnachie2012,
  author  = {{McConnachie}, Alan W.},
  title   = {The Observed Properties of Dwarf Galaxies in and around the Local Group},
  journal = {AJ},
  year    = {2012},
  month   = {July},
  volume  = {144},
  number  = {1},
  eid     = {4},
  pages   = {4},
  doi     = {10.1088/0004-6256/144/1/4},
  eprint  = {1204.1562},
  archivePrefix = {arXiv},
  primaryClass  = {astro-ph.CO}
}

@ARTICLE{eskew2012,
  author  = {{Eskew}, Michael and {Zaritsky}, Dennis and {Meidt}, Sharon},
  title   = {Converting from 3.6 and 4.5 Micron Fluxes to Stellar Mass},
  journal = {AJ},
  year    = {2012},
  month   = {June},
  volume  = {143},
  number  = {6},
  eid     = {139},
  pages   = {139},
  doi     = {10.1088/0004-6256/143/6/139}
}

@ARTICLE{skibba2012,
  author  = {{Skibba}, Ramin A. and {Engelbracht}, Charles W. and
             {Aniano}, Gonzalo and others},
  title   = {The Spatial Distribution of Dust and Stellar Emission of the Magellanic Clouds},
  journal = {ApJ},
  year    = {2012},
  month   = {December},
  volume  = {761},
  number  = {1},
  eid     = {42},
  pages   = {42},
  doi     = {10.1088/0004-637X/761/1/42}
}

\end{document}